# On the evolution of fuel droplet evaporation zone and its interaction with the flame front in ignition of spray flames

Qiang Li, Chang Shu, Huangwei Zhang[*]

*Department of Mechanical Engineering, National University of Singapore, 9 Engineering Drive 1, Singapore 117576, Republic of Singapore*

**Abstract**

Evolution of fuel droplet evaporation zone and its interaction with the propagating flame front are studied in this work. A general theory is developed to describe the evolutions of flame propagation speed, flame temperature, droplet evaporation onset and completion locations in ignition and propagation of spherical flames. The influences of liquid droplet mass loading, heat exchange coefficient (or evaporation rate) and Lewis number on spherical spray flame ignition are studied. Two flame regimes are considered, i.e., heterogeneous and homogeneous flames, based on the mixture condition near the flame front. The results indicate that the spray flame trajectories are considerably affected by the ignition energy addition. The critical condition for successful ignition for the fuel-rich mixture is coincidence of inner and outer flame balls from igniting kernel and propagating flame. The flame balls always exist in homogeneous mixtures, indicating that ignition failure and critical successful events occur only in purely gaseous mixture. The fuel droplets have limited effects on minimum ignition energy, which however increases monotonically with the Lewis number. Moreover, flame kernel originates from heterogeneous mixtures due to the initially dispersed droplets near the spark. The evaporative heat loss in the burned and unburned zones of homogeneous and heterogeneous spray flames is also evaluated, and the results show that for the failed flame kernels, evaporative heat loss behind and before the flame front first increases and then decreases. The evaporative heat loss before the flame front generally increases, although non-monotonicity exists, when the flame is successfully ignited and propagate outwardly. For heterogeneous flames, the ratio of the heat loss from the burned zone to the total one decreases as the flame expands. Moreover, droplet mass loading and heat exchange coefficient considerably affect the evaporating heat loss from burned and unburned zones.

---

[*] Corresponding author. Email: huangwei.zhang@nus.edu.sg, Tel: +65 6516 2557.

# 1   Introduction

Successful flame ignition in sprays is important in various combustion systems with liquid fuels, such as aero engines, internal combustion engines, and rocket engines. Compared to gaseous flames, existence of liquid fuel sprays may bring intriguing features in the ignition process [1–3]. In practical combustion devices, the flame is normally initiated by an electrical spark [4] or laser beam [5] with external energy deposition. Mastorakos [1] identifies three stages of spray flame ignition: (1) kernel generation, (2) flame growth, and (3) burner-scale flame establishment. In each stage, the flame is considerably affected by liquid fuel droplet characteristics, including droplet spatial distribution, movement or dispersion, heating, and evaporation. These influences are realized through comprehensive interphase momentum, mass, and heat exchanges.

Dynamic evolution of the gas−liquid mixtures near the flame front may result in two distinctive cases of spray flame ignition and expansion: (I) fuel sprays only exist only ahead of the flame front due to fast heating and evaporation; (II) fuel sprays exist in both pre- and post-flame areas. The occurrence of these two scenarios is affected by liquid fuel, sprayed droplets, and/or gaseous flame properties, and both are indeed observed from spray flame ignition experiments. For instance, Akamatsu et al. found that remaining droplets are distributed behind the nonluminous flame, which continue burning randomly and discontinuously, leading to a luminous flame in the product gas [6]. This two-layer structure of spray flames is also recorded by Fan et al. [7]. However, Atzler et al. took the laser sheet image of a laminar iso-octane aerosol flame and found that the fuel droplets are completely vaporized before the flame front, resulting in purely gaseous mixture in the burned zone [8]. More recently, de Oliveira et al. studied the droplet dynamics in spray flame ignition and propagation in a more microscopic (droplet scale) way with the aid of the advanced optical measurements [5,9]. They identified various flame propagation modes based on OH* and OH (hydroxyl) planar laser-induced fluorescence images, from which the unsteady process of droplet crossing the flame front, evaporating/burning in the burned area can be seen clearly. However, the above work did not provide explanations about the conditions under which the scenarios I and/or II can exist and how they evolve in a highly transient combustion process, e.g., spark ignition.

Detailed numerical simulations provide us further insights about the droplet−flame interactions in the scenarios I and/or II mentioned above. For example, Wandel et al. [10] studied turbulent spray flame ignition and their results show that droplet evaporation and distribution are of great important to initiate a flame kernel. Also, Ozel et al. [11] found that fuel droplets can reside or diminish in the burned zone for both turbulent and laminar flame kernels and the droplet−flame interaction considerably modulate the flame curvature. Neophytou et al. [12] studied the spark ignition and edge flame propagation in mixing layers and their results show that the larger kernel can be achieved when the spark is close to the sprays, whilst the ignition is successful even if the spark is in the oxidizer side, as along as the droplets can evaporate rapidly. Furthermore, Thimothée et al. [13] studied the passage of liquid droplets through a spherical flame front and it is seen that droplet size and inter-droplet distance are the most important controlling factor.

In the above research efforts, specific fuels, droplet sizes, and flow configurations are considered, and therefore the observations may lack of generality. Further studies are needed for developing a general theory for describing the droplet-flame interaction in spray flame ignition. There has been some theoretical analysis available, which gives us coherent overall picture about spray flame dynamics. Greenberg [14] derived an evolution equation for a laminar flame propagation into fuel sprays, considering finite-rate evaporation and droplet drag effects. With the similar model, Han and Chen [15] further examined the influences of finite-rate evaporation on spray flame propagation and ignition, and find that flame propagation speed, Markstein length and minimum ignition energy are strongly affected by droplet loading and evaporation rate. Nonetheless, in their work [14,15], the droplets are assumed to be always distributed in the full domain, and hence the unsteady evolutions of droplet distribution are not considered. Recently, Li et al. [16,17] studied propagation of planar and spherical flames in fuel droplet mists with evolving droplet distributions considered. It is found that the mixtures (gas−only or vapour/droplets) around the flame front is critical for flame propagation due to the evaporative heat loss and fuel vapor availability. Nevertheless, the general mechanisms about how dynamic evolutions of droplet distribution affects flame ignition are not discussed and hence still not clear.

In this work, we aim to theoretically study the interactions between evaporating droplets and spray flame in ignition of partially pre-vaporized fuel sprays. Dynamic droplet distributions caused by droplet evaporation and its interactions with an evolving spherical flame are incorporated in our theoretical model, through introducing the locations for onset and completion of droplet evaporation [16,17]. The influences of liquid fuel and gas properties on the spherical flame ignition will be examined, including evaporation heat exchange coefficient (or evaporation rate), droplet mass loading, and Lewis number. The rest of the paper is structured as below. The physical and mathematical models are presented in Section 2, whilst the analytical solutions are listed in Section 3. The results will be discussed in Section 4, and Section 5 closes the paper with conclusions.

## 2 Physical and mathematical model

### 2.1 Physical model

Spark ignition of one-dimensional spherical flame in partially pre-vaporized fuel sprays will be studied in this study. Initially, the gaseous fresh mixture is assumed to be fuel-rich and the dilute fuel droplets are uniformly dispersed. The ignition is modelled as a localized energy deposition in fuel sprays at the spherical center. A spray flame kernel is generated and continuously propagate outwardly, if the ignition energy is larger than the minimum ignition energy. During the flame development process, based on the droplet distribution relative to the reaction front, two general scenarios exist as mentioned in the last section, i.e., evaporating droplets in (I) both pre- and post-flame zones and (II) pre-flame zone only.

The sketches of the two physical models are shown in Figs. 1(a) and 1(b), respectively. There are three characteristic locations for liquid and gas phases, including flame front ($R_f$), droplet evaporation onset ($R_v$) and completion ($R_c$) fronts. $R_v$ corresponds to the location where the droplets are just heated up to boiling temperature and hence start to evaporate. For $R < R_v$, the droplet temperature remains constant and evaporation continues [15,18–21]. The evaporation onset front $R_v$ is always before the flame front $R_f$, indicating that onset of droplet vaporization spatially precedes the gaseous combustion. Moreover, $R_c$ denotes the location at which all the droplets are critically vaporized. Therefore, for $R < R_c$, no droplets are left and hence their effects on the gas phase diminish. In this sense, the droplet completion front is a two-phase contact surface in nature, which spatially demarcates the purely gaseous and liquid−gas two phase mixtures.

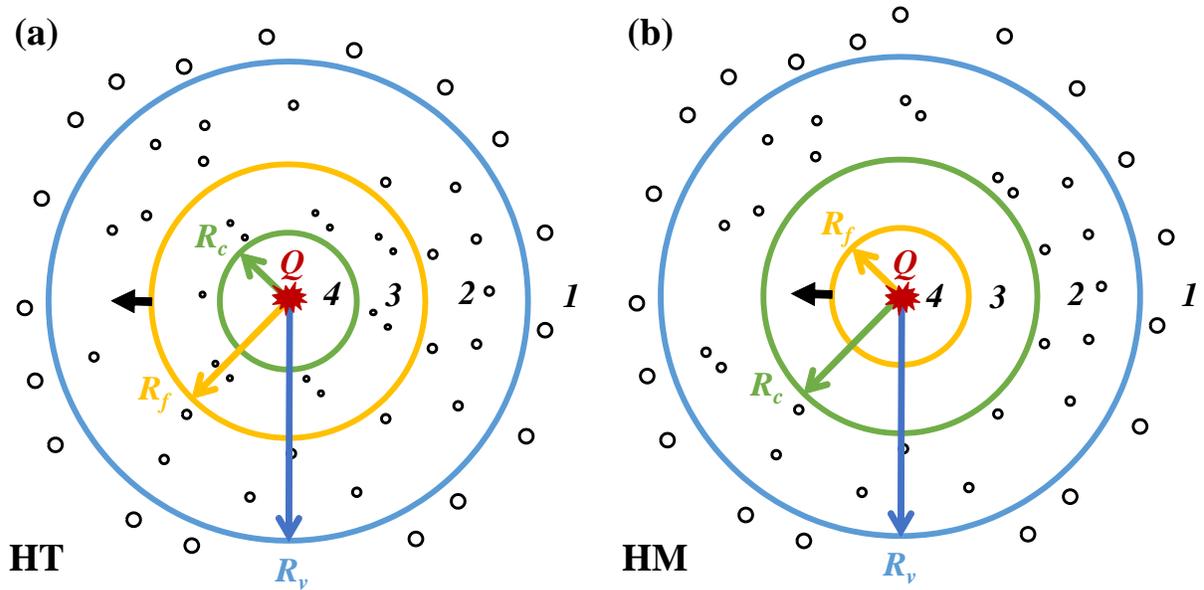

Fig.1 Schematic of outwardly propagating spherical flames in liquid fuel sprays: (a) heterogeneous flame, (b) homogeneous flame. Circle: fuel droplet. Yellow line: flame front ($R_f$); green line: evaporation completion front ($R_c$); blue line: evaporation onset front ($R_v$). Black arrow: flame propagation direction. Red Spark ($Q$): forced ignition energy. Numbers indicate the combustion and droplet evaporation zones.

In Fig. 1(a), the evaporation completion front $R_c$ lies behind the flame front (i.e., $R_c < R_f < R_v$). Therefore, the mixture around the propagating flame front $R_f$ is composed of gaseous vapor and evaporating fuel droplets. In Fig. 1(b), it is located before the flame front (i.e., $R_f < R_c < R_v$), and the mixture around the flame front is purely gaseous since all the droplets have been gasified into vapor in the preheated area. For brevity, hereafter, we term the first (Fig. 1a) and second (Fig. 1b) cases as *heterogeneous* (abbreviated as "HT") and *homogeneous* ("HM") *flames*, respectively. There are four zones in both flames, as shown in Fig. 1. Specifically, zone 1 represents the pre-vaporization zone before $R_v$, and 2 indicates pre-flame evaporation zone before $R_f$ for heterogeneous flame and before $R_c$ for homogeneous flame. As for zone 3, it represents post-flame evaporation zone before $R_c$ for heterogeneous flame, and pre-flame zone without evaporation for homogeneous flame. Meanwhile, zone 4 is the post-flame droplet-free zone for both flames.

In the following, we will develop a general theory to describe ignition and propagation of premixed spray flames, considering evolving droplet evaporation zones with propagating reaction front and transition between homogeneous and heterogeneous flames. It should be noted that this model is different from the finite-rate evaporation model used by Han and Chen [15] and Greenberg [14]. The novelty of our model is introduction of the evolving two-phase contact surface, which can provide greater flexibilities in describing the interactions between the expanding spray flame and the evaporating fuel droplets.

## 2.2 Governing equation

For the gaseous flames, the diffusive-thermal model [22,23] is adopted, with which the thermal and transport properties (e.g. density, thermal conductivity, and heat capacity) are assumed to be constant and the convection flux is absent. This model has been used in numerous studies on flame dynamics with both gaseous and liquid fuels [15,18,20,21,24,25] and its applicability has been confirmed in the previous work. One-step chemistry is considered, i.e., $F + O \rightarrow P$, with $F$, $O$ and $P$ being fuel vapour, oxidizer and product, respectively. Only fuel-rich vapour/air mixture (i.e., equivalence ratios of fuel vapor are above unity) is studied in this work and oxidizer $O$ is the deficient species. Therefore, the equations for gas temperature and oxidizer mass fraction are

$$\tilde{\rho}_g \tilde{C}_{p,g} \frac{\partial \tilde{T}}{\partial \tilde{t}} = \frac{1}{\tilde{r}^2} \frac{\partial}{\partial \tilde{r}} \left( \tilde{r}^2 \tilde{\lambda}_g \frac{\partial \tilde{T}}{\partial \tilde{r}} \right) + \tilde{q}_c \tilde{\omega}_c - \tilde{q}_v \tilde{\omega}_v, \tag{1}$$

$$\tilde{\rho}_g \frac{\partial \tilde{Y}_O}{\partial \tilde{t}} = \frac{1}{\tilde{r}^2} \frac{\partial}{\partial \tilde{r}} \left( \tilde{r}^2 \tilde{\rho}_g \tilde{D}_O \frac{\partial \tilde{Y}_O}{\partial \tilde{r}} \right) - \tilde{\omega}_c, \tag{2}$$

where the tilde symbol ~ is used to indicate that the variables are dimensional. $\tilde{t}$ and $\tilde{r}$ are respectively the temporal and spatial coordinates. $\tilde{T}$, $\tilde{\rho}_g$, $\tilde{C}_{p,g}$, and $\tilde{\lambda}_g$ are the gas temperature, density, heat capacity and thermal conductivity, respectively. $\tilde{Y}_O$ and $\tilde{D}_O$ are the mass fraction and molecular diffusivity of the oxidizer. $\tilde{q}_c$ is the reaction heat release per unit mass of the oxidizer. $\tilde{q}_v$ is the latent heat of vaporization of liquid fuel, whilst $\tilde{\omega}_v$ is the evaporation rate of the fuel droplet. The chemical reaction rate $\tilde{\omega}_c$ in Eq. (1) reads

$$\tilde{\omega}_c = \tilde{\rho}_g \tilde{A} \tilde{Y}_o \, exp(-\tilde{E}/\tilde{R}^0 \tilde{T}). \tag{3}$$

Here $\tilde{A}$ is the pre-exponential factor, $\tilde{E}$ is the activation energy for the reaction, and $\tilde{R}^0$ is the universal gas constant.

For liquid fuel droplets, monodispersed dilute sprays in a pre-vaporized fuel/air mixture are considered. The droplets are spherical and their properties (e.g., density and heat capacity) are assumed to be constant and are uniformly distributed initially. Due to the dilute droplet concentration, inter-droplet collisions are not considered and therefore diffusion of liquid droplets can be neglected. Besides, kinematic equilibrium is assumed between gas and fuel droplets. The validity of the above assumptions is confirmed in previous theoretical work on two-phase flames [26–29]. Furthermore, in zone 1 (pre-vaporization, see Fig. 1), interphase thermal equilibrium is assumed, and hence the two phases have the same temperature [15,18–21]. The Eulerian description is adopted for the liquid phase and hence the equation for droplet mass loading $Y_d$ ($\equiv \tilde{N}_d \tilde{m}_d / \tilde{\rho}_g$, $\tilde{N}_d$ is droplet number density) reads

$$\frac{\partial Y_d}{\partial \tilde{t}} = -\frac{\tilde{\omega}_v}{\tilde{\rho}_g}. \tag{4}$$

We assume that the heat transferred from the surrounding gas to the droplets is completely used for phase change of liquid fuel, which is related to the latent heat of evaporation $\tilde{q}_v$ [18,19,21,30]. Therefore, $\widetilde{\omega}_v$ in Eqs. (1), (2), and (4) can be modelled as

$$\widetilde{\omega}_v = \frac{\widetilde{N}_d \tilde{s}_d \tilde{h}(\tilde{T}-\tilde{T}_v)H(\tilde{T}-\tilde{T}_v)}{\tilde{q}_v}, \tag{5}$$

where $\tilde{s}_d = \pi \tilde{d}^2$ is the surface area of a single droplet, $\tilde{d}$ is the droplet diameter, $Nu$ is the Nusselt number, $\tilde{T}_v$ is the boiling temperature of the liquid fuel. $H(\tilde{T} - \tilde{T}_v)$ is the Heaviside function, which indicates that the evaporation only occurs when the local temperature is above the boiling temperature. $\tilde{h}$ is the heat exchange coefficient, estimated using the Ranz and Marshall correlation [31]

$$Nu = \frac{\tilde{h}\tilde{d}}{\tilde{\lambda}_g} = 2.0 + 0.6\, Re^{1/2}\, Pr^{1/3}, \tag{6}$$

where $Nu$, $Pr$ and $Re$ are the Nusselt number, Prandtl number and droplet Reynolds number, respectively. We can neglect the effect of droplet Reynolds number due to the assumption of kinematic equilibrium and therefore $Nu \approx 2$. Accordingly, the evaporation rate $\widetilde{\omega}_v$ can be re-written as

$$\widetilde{\omega}_v = \widetilde{N}_d \tilde{s}_d \tilde{\lambda}_g Nu(\tilde{T} - \tilde{T}_v)H(\tilde{T} - \tilde{T}_v)/(\tilde{d}\tilde{q}_v). \tag{7}$$

To render the analytical analysis more general, normalization of Eqs. (1), (2) and (4) can be performed, with the following reference parameters for the velocity, length, time, mass fraction and temperature

$$u = \frac{\tilde{u}}{\tilde{u}_b}, r = \frac{\tilde{r}}{\tilde{l}_{th}}, t = \frac{\tilde{t}}{\frac{\tilde{l}_{th}}{\tilde{u}_b}}, Y = \frac{\tilde{Y}}{\tilde{Y}_0}, T = \frac{\tilde{T}-\tilde{T}_0}{\tilde{T}_b-\tilde{T}_0}. \tag{8}$$

Here $\tilde{T}_0$ and $\tilde{Y}_0$ denote the temperature and fuel mass fraction of the pre-vaporized mixture, respectively. $\tilde{u}_b$, $\tilde{T}_b = \tilde{T}_0 + \tilde{q}_c\tilde{Y}_0/\tilde{C}_{p,g}$ and $\tilde{l}_{th} = \widetilde{D}_{th}/\tilde{u}_b$ are respectively the laminar flame speed, adiabatic flame temperature and flame thickness of the pre-vaporized gas mixture. $\widetilde{D}_{th} = \tilde{\lambda}_g/\tilde{\rho}_g\tilde{C}_{p,g}$ is the thermal diffusivity.

Following previous theoretical analysis for both gaseous flames and two-phase flames with dispersed liquid droplets [15,20,24,32–37], we adopt the quasi-steady state assumption in the moving coordinate system attached to the propagating flame front $R_f(t)$, i.e. $\eta = r - R_f(t)$. This assumption has been extensively validated by transient numerical simulations for gaseous spherical flames [24,32–34,38], in which the unsteady effects are found to have a limited influence based on the budget analysis of diffusion, reaction and convection terms in propagating spherical flames. Moreover, due to relatively dilute fuel droplet concentration, their effects on the reaction zone thickness are small and therefore gaseous combustion still dominates [15,19]. Besides, due to the kinematic equilibrium between two phases, the droplets approximately follow the carrier gas. Therefore, the non-dimensional equations of Eqs. (1), (2) and (4) can be respectively written as

$$-U\frac{dT}{d\eta} = \frac{1}{(\eta+R_f)^2}\frac{d}{d\eta}\left[(\eta+R_f)^2\frac{dT}{d\eta}\right] + \omega_c - q_v\omega_v, \tag{9}$$

$$-U\frac{dY_O}{d\eta} = \frac{Le^{-1}}{(\eta+R_f)^2}\frac{d}{d\eta}\left[(\eta+R_f)^2\frac{dY_O}{d\eta}\right] - \omega_c, \tag{10}$$

$$-U\frac{dY_d}{d\eta} = -\omega_v, \tag{11}$$

where $U = dR_f/dt$ is the non-dimensional flame propagating speed. $q_v = \tilde{q}_v/[\tilde{C}_{p,g}(\tilde{T}_b - \tilde{T}_0)]$ is the normalized latent heat of vaporization. $Le = \tilde{D}_{th}/\tilde{D}_O$ is the Lewis number of the deficient species, i.e., oxidizer. The normalized chemical reaction rate $\omega_c$ reads

$$\omega_c = \frac{1}{2Le} Y_O Z^2 \exp\left[\frac{Z(T-1)}{\sigma+(1-\sigma)T}\right], \qquad (12)$$

where $Z$ is the Zel'dovich number and $\sigma$ is the thermal expansion ratio. It is assumed that $Z = 10$ and $\sigma = 0.15$, respectively, following Refs. [15,18,20,21,24,25], which correspond to the properties of typical hydrocarbon fuels.

The term $\omega_v$ in Eqs. (9) and (11) is the non-dimensional droplet evaporation rate, i.e.,

$$\omega_v = \frac{\Omega(T-T_v)}{q_v} H(T - T_v), \qquad (13)$$

where $T_v$ is the non-dimensional boiling temperature and assumed to be $T_v = 0.15$ [15,39]. The normalized latent heat of vaporization $q_v$ is 0.4. The foregoing parameters are close to those of typical liquid hydrocarbon fuels (e.g., ethanol) [39]. The heat exchange coefficient $\Omega$ is

$$\Omega = \pi \tilde{N}_d Nu \tilde{d} \tilde{D}_{th}^2 \tilde{u}_b^{-2}. \qquad (14)$$

As shown in Eq. (14), $\Omega$ is a lumped parameter with both gas and droplet properties [18,19]. The heat exchange coefficient $\Omega$ is a measure of convective heat transfer between the gas and liquid droplets, and it is affected by the droplet diameter and number density. Moreover, since the latent heat $q_v$ is fixed in this study, higher $\Omega$ generally indicates faster droplet evaporation rate $\omega_v$, as seen from Eq. (13). To avoid the nonlinearity in Eq. (11), the weak dependence of $\Omega$ on $Y_d$ ($\Omega \sim Y_d^{1/3}$) is not considered, following Belyakov et al. [19].

## 2.3 Jump and boundary conditions

The non-dimensional boundary conditions for both gaseous phase ($T$ and $Y_O$) and liquid phase ($Y_d$) at the left boundary (spherical center, $\eta = -R_f$) and the right boundary ($\eta \to +\infty$) are [15,18–21]

$$\eta = -R_f: (\eta + R_f)^2 \frac{dT}{d\eta} = -Q, \quad \frac{dY_O}{d\eta} = 0, \quad Y_d = 0, \tag{15}$$

$$\eta \to +\infty: T = 0, \quad Y_O = 1, \quad Y_d = \delta. \tag{16}$$

Here $\delta$ is the initial mass loading of the fuel droplet in the fresh mixture, and $Q$ is the normalized ignition energy.

At the evaporation onset front, $\eta = \eta_v$, the gas temperature ($T$), oxidizer mass fraction ($Y_O$), and fuel droplet mass loading ($Y_d$) satisfy the following jump conditions [15,18–21]

$$T = T_v, \quad [T] = \left[\frac{dT}{d\eta}\right] = [Y_o] = \left[\frac{dY_o}{d\eta}\right] = 0, Y_d = \delta. \tag{17}$$

where the square brackets, i.e., $[f] = f(\eta^+) - f(\eta^-)$, denote the difference between the variables at two sides of a location.

At the evaporation completion front, $\eta = \eta_c$, the jump conditions for the gas temperature ($T$), oxidizer mass fraction ($Y_O$), and droplet mass loading ($Y_d$) take the following form [19]

$$\begin{cases} [T] = \left[\frac{dT}{d\eta}\right] = [Y_o] = \left[\frac{dY_o}{d\eta}\right] = 0, Y_d = 0, & \eta_c > 0 \\ [T] = \left[\frac{dT}{d\eta}\right] = [Y_o] = \left[\frac{dY_o}{d\eta}\right] = 0, [Y_d] = 0, -R_f < \eta_c < 0 \end{cases}. \tag{18}$$

It should be noted that the condition of $-R_f < \eta_c$ ($\eta_c = -R_f$ is excluded) indicates the evaporation completion front is formed near or well ahead of the spherical center. Physically, when the external energy is initially deposited in fully dispersed fuel mists, it would take a finitely long time to heat and gasify the droplets near the spark. Therefore, during the flame ignition process, there would be a finitely long period when the droplets are still fully distributed, although they are continuously evaporating. This corresponds to $\eta_c = -R_f$, with which our theory is mathematically intractable due to the inconsistency of the jump ($\eta = \eta_c$) and boundary ($\eta = -R_f$) conditions at the spherical center. Therefore, the starting instant our theory can describe is when the evaporation completion front critically deviates from the spark location, i.e., $\eta_c > -R_f$.

Large activation energy of the gas phase reaction is assumed [22]. Its validity has been confirmed in numerous theoretical analysis of both gaseous [24,32,33,38,40,41] and particle- or droplet-laden [14,15,21,25–28] flames. It is adequate to predict the main flame dynamics, such as ignition and propagation. With this assumption, chemical reaction is confined at an infinitesimally thin sheet (i.e., $\eta = 0$). The corresponding jump conditions are

$$T = T_f, \ Y_o = [Y_d] = 0, \tag{19}$$

$$-\left[\frac{dT}{d\eta}\right] = \frac{1}{Le}\left[\frac{dY_o}{d\eta}\right] = [\sigma + (1-\sigma)T_f]^2 \exp\left[\frac{Z}{2}\left(\frac{T_f-1}{\sigma+(1-\sigma)T_f}\right)\right], \tag{20}$$

where $T_f$ is the flame temperature.

## 3 Analytical solution

The governing equations (9) – (11) with proper boundary and jump conditions (i.e., Eqs. 15 – 19) can be solved analytically. The solutions for gas temperature $T$, oxidizer mass fraction $Y_o$, and droplet mass loading $Y_d$ in zones 1-4 of both homogeneous and heterogeneous flames are presented in Section 3.1. Moreover, the correlations describing flame speed $U$, flame temperature $T_f$, evaporation onset location $\eta_v$ and droplet completion location $\eta_c$ under different flame radii $R_f$ are also derived in Section 3.2.

## 3.1 Solutions of $T$, $Y_O$ and $Y_d$

For heterogeneous flame, the solutions of gas temperature $T$, oxidizer mass fraction $Y_O$, and droplet loading $Y_d$ in zones 1 to 4 are (the number subscripts indicate different zones as shown in Fig. 1)

$$\begin{cases} T_1(\eta) = T_v \frac{I(\eta,U)}{I(\eta_v,U)} \\ T_2(\eta) = T_v + k_1 L_1(\eta) + k_2 L_2(\eta) \\ T_3(\eta) = T_v + \zeta_1 L_1(\eta) + \zeta_2 L_2(\eta) \\ T_4(\eta) = Q[I(\eta_c, U) - I(\eta, U)] + T_v + \zeta_1 L_1(\eta_c) + \zeta_2 L_2(\eta_c) \end{cases}, \quad (21)$$

$$\begin{cases} Y_{O,1}(\eta) = 1 - \frac{I(\eta, LeU)}{I(0, LeU)} \\ Y_{O,2}(\eta) = 1 - \frac{I(\eta, LeU)}{I(0, LeU)}, \\ Y_{O,3}(\eta) = 0 \\ Y_{O,4}(\eta) = 0 \end{cases} \quad (22)$$

$$\begin{cases} Y_{d,1}(\eta) = \delta \\ Y_{d,2}(\eta) = \delta + \frac{\Omega}{Uq_v} \int_{\eta_v}^{\eta} [T_2(\eta) - T_v] d\eta \\ Y_{d,3}(\eta) = \frac{\Omega}{Uq_v} \int_{\eta_c}^{\eta} [T_3(\eta) - T_v] d\eta \\ Y_{d,4}(\eta) = 0 \end{cases}. \quad (23)$$

Here $I(x, y)$, $L_1(\eta)$ and $L_2(\eta)$ are

$$I(x, y) = e^{-yR_f} \int_x^{+\infty} (\xi + R_f)^{-2} e^{-y\xi} d\xi, \quad (24)$$

$$L_1(\eta) = exp\left[-\frac{(K+U)(\eta+R_f)}{2}\right] M\left(1 + \frac{U}{K}, 2, K(\eta + R_f)\right), \quad (25)$$

$$L_2(\eta) = exp\left[-\frac{(K+U)(\eta+R_f)}{2}\right] N\left(1+\frac{U}{K}, 2, K(\eta+R_f)\right). \tag{26}$$

Here $K = \sqrt{U^2 + 4\Omega}$. $M(a,b,c)$ and $N(a,b,c)$ are the *Kummer confluent hypergeometric function* and the *Tricomi confluent hypergeometric function*, respectively [42]. They can be expressed as

$$M(a,b,c) = \frac{\Gamma(b)}{\Gamma(b-a)\Gamma(a)} \int_0^1 e^{ct} t^{a-1}(1-t)^{b-a-1} dt, \tag{27}$$

$$N(a,b,c) = \frac{1}{\Gamma(a)} \int_0^{+\infty} t^{a-1} e^{-ct}(1+t)^{b-a-1} d\xi, \tag{28}$$

where $\Gamma(x) = \int_0^{+\infty} t^{x-1} e^{-t} dt$.

Moreover, the expressions for $k_1$, $k_2$, $\zeta_1$, and $\zeta_2$ in Eq. (21) are respectively

$$k_{1,2} = T_v \frac{I'(\eta_v, U)}{I(\eta_v, U)} \frac{L_{2,1}(\eta_v)}{L'_{1,2}(\eta_v) L_{2,1}(\eta_v) - L'_{2,1}(\eta_v) L_{1,2}(\eta_v)}, \tag{29}$$

$$\zeta_{1,2} = \frac{(T_f - T_v) L'_{2,1}(\eta_c) + Q L_{2,1}(0) I'(\eta_c, U)}{L'_{2,1}(\eta_c) L_{1,2}(0) - L'_{1,2}(\eta_c) L_{2,1}(0)}, \tag{30}$$

where $I'(x,y) = \frac{\partial I(x,y)}{\partial x} = -e^{-y(x+R_f)}(x+R_f)^{-2}$.

For homogeneous flame, the solutions of $T$, $Y_o$, and $Y_d$ in zones 1 to 4 are

$$\begin{cases} T_1(\eta) = T_v \frac{I(\eta, U)}{I(\eta_v, U)} \\ T_2(\eta) = T_v + k_1 L_1(\eta) + k_2 L_2(\eta) \\ T_3(\eta) = T_v + k_1 L_1(\eta_c) + k_2 L_2(\eta_c) + \frac{k_1 L'_1(\eta_c) + k_2 L'_2(\eta_c)}{DI(\eta_c, U)} [I(\eta, U) - I(\eta_c, U)] \\ T_4(\eta) = Q[I(0, U) - I(\eta, U)] + T_f \end{cases}, \tag{31}$$

$$\begin{cases} Y_{O,1}(\eta) = 1 - \frac{I(\eta, LeU)}{I(0, LeU)} \\ Y_{O,2}(\eta) = 1 - \frac{I(\eta, LeU)}{I(0, LeU)} \\ Y_{O,3}(\eta) = 1 - \frac{I(\eta, LeU)}{I(0, LeU)} \\ Y_{O,4}(\eta) = 0 \end{cases}, \qquad (32)$$

$$\begin{cases} Y_{d,1} = \delta \\ Y_{d,2} = \delta + \frac{\Omega}{Uq_v} \int_{\eta_v}^{\eta} (T_2 - T_v) d\eta \\ Y_{d,3} = 0 \\ Y_{d,4} = 0 \end{cases}. \qquad (33)$$

## 3.2 Correlations for spherical flame and fuel sprays

The correlations between flame radius $R_f$, flame propagation speed $U$, flame temperature $T_f$, and droplet characteristic locations, $\eta_v$ and $\eta_c$, can be derived through the jump conditions (i.e., Eqs. 19–20) in Section 2.3. If the reactants around the flame front $R_f$ are heterogeneous, then the following correlations hold

$$(\zeta_1 - k_1)L_1'(0) + (\zeta_2 - k_2)L_2'(0) = [\sigma + (1-\sigma)T_f]^2 \exp\left[\frac{Z}{2}\left(\frac{T_f - 1}{\sigma + (1-\sigma)T_f}\right)\right], \qquad (34)$$

$$\frac{1}{Le} \frac{R_f^{-2} e^{-LeUR_f}}{I(0, LeU)} = [\sigma + (1-\sigma)T_f]^2 \exp\left[\frac{Z}{2}\left(\frac{T_f - 1}{\sigma + (1-\sigma)T_f}\right)\right], \qquad (35)$$

$$T_v + k_1 L_1(0) + k_2 L_2(0) = T_f, \qquad (36)$$

$$\delta + \frac{\Omega}{Uq_v} \int_{\eta_v}^{0} (T_2 - T_v) d\eta = \frac{\Omega}{Uq_v} \int_{\eta_c}^{0} (T_3 - T_v) d\eta. \qquad (37)$$

If the reactants around the flame front $R_f$ are homogeneous, then the correlations are

$$-\left[Q + \frac{k_1 L_1'(\eta_c) + k_2 L_2'(\eta_c)}{DI(\eta_c, U)}\right] R_f^{-2} e^{-UR_f} = [\sigma + (1-\sigma)T_f]^2 \exp\left[\frac{Z}{2}\left(\frac{T_f - 1}{\sigma + (1-\sigma)T_f}\right)\right], \qquad (38)$$

$$\frac{1}{Le}\frac{R_f^{-2}e^{-LeUR_f}}{I(0,LeU)} = [\sigma + (1-\sigma)T_f]^2 \exp\left[\frac{Z}{2}\left(\frac{T_f-1}{\sigma+(1-\sigma)T_f}\right)\right], \tag{39}$$

$$T_v + k_1 L_1(\eta_c) + k_2 L_2(\eta_c) + \frac{k_1 L_1'(\eta_c) + k_2 L_2'(\eta_c)}{DI(\eta_c,U)}[I(0,U) - I(\eta_c,U)] = T_f, \tag{40}$$

$$\delta + \frac{\Omega}{Uq_v}\int_{\eta_v}^{\eta_c}(T_2 - T_v)d\eta = 0. \tag{41}$$

The implications of the four equations in the correlations for heterogeneous and homogeneous flames are:

a) *Equations (34) and (38)*: heat absorbed by the gaseous mixture and droplet is equal to the heat produced by chemical reaction.

b) *Equations (35) and (39)*: energy from the initial oxidizer vapor is equal to the heat from chemical reaction.

c) *Equations (36) and (40)*: continuity of gas temperature at the flame front for the pre-flame evaporation zone (zone 2).

d) *Equation (37)*: continuity of droplet loading at the flame front for the pre- and post-flame evaporation zones in heterogeneous flames.

e) *Equation (41)*: continuity of droplet loading at the evaporation completion front ($Y_d(\eta_c) = 0$) for the pre-flame evaporation zone in homogeneous flames.

The open-source GNU Scientific Library (GSL) [43] is used to numerically solve the above non-linear systems, i.e., Eqs. (34)−(37) and (38)−(41). Therefore, ignition of premixed flames in pre-vaporized liquid fuel sprays, interactions between the gas and liquid phases, and transition between homogeneous and heterogeneous flames will be analysed in Section 4, through the evolutions of flame propagation speed, flame temperature, and characteristic locations of droplet evaporation.

## 4 Results and discussion

### 4.1 Spray flame ignition

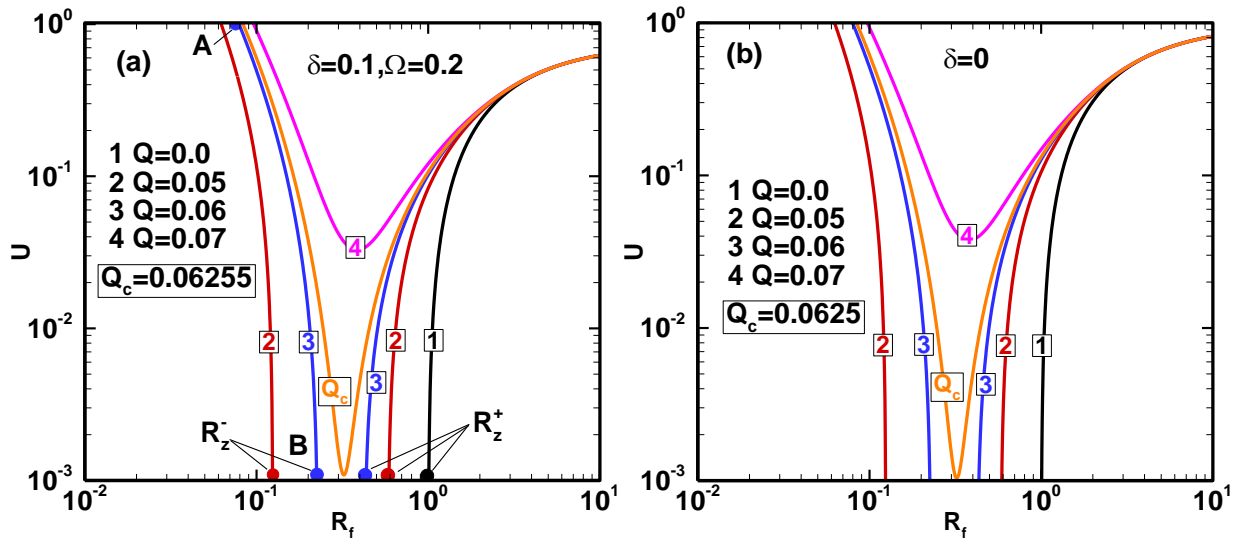

Fig. 2 Flame propagation speed as a function of flame radius with different ignition energies: (a) $\delta = 0.1$, $\Omega = 0.2$ and (b) $\delta = 0$ (gaseous flame). $Le = 1.0$, $R_Z^+$ and $R_Z^-$: flame ball radius.

Figure 2(a) shows the flame propagation speed $U$ as a function of flame radius $R_f$ at different ignition energies $Q$ in liquid fuel sprays of mass loading $\delta = 0.1$ and heat exchange coefficient $\Omega = 0.2$. The Lewis number of the gaseous mixtures is 1.0. It is seen that the propagation trajectories of the spray flames are considerably affected by the ignition energy. Specifically, when $Q = 0$ (line #1), the spray flame is initiated at a stationary flame ball (termed as outer flame ball, with its radius marked as $R_Z^+$) with a radius of about 1.0 and propagate outwardly towards spherical flame with lager radii. This $U - R_f$ curve is deemed flame propagating branch. When the external ignition energy is deposited (e.g., $Q = 0.05$, line #2), a new left branch with smaller flame radii emerges, which corresponds to flame kernel branch. This branch starts at a high propagation speed ($U > 1$), resulting from the overdrive effects by the ignition energy. However, the flame kernel propagating speed decays quickly until a stationary flame ball is formed ($U = 0$, inner flame ball, marked with $R_Z^-$ in Fig. 2a), with which the flame propagation in fuel sprays is terminated. This situation corresponds to ignition failure. de Oliveria et al. also reported the similar failure mode for spray flame ignition (so-called long-term mode), in which the flame propagation is terminated at some radii [5]. With further increased ignition energy (e.g., line #3), the flame kernel branch and the flame propagating branch move towards each other, with gradually approaching flame ball radii, $R_Z^-$ and $R_Z^+$. When the ignition energy is equal to the MIE $Q_c$ (e.g., 0.06255 in Fig. 2a), the twin flame ball solutions coincide, i.e., $R_Z^+ = R_Z^-$, and therefore the two flame branches merge, which implies that the flame can critically evolve from igniting kernel to stably propagation state. The corresponding flame radius is critical ignition radius, i.e., $R_{ic} = R_Z^+ = R_Z^-$. Similar critical conditions controlled by flame ball evolutions are also achieved for ignition of gaseous and spray flames [32,33,44–46]. For $Q > Q_c$ (e.g., line #4), the flame kernel grows continuously, propagate outwardly, and eventually reach the spherical flames with large radii.

The gaseous flame ignition in vapour-rich and droplet-free mixture is also shown in Fig. 2(b) for comparison. We can see that the flame trajectories in spray flames and gaseous flames for fixed Lewis number are qualitatively similar, indicating that the dilute fuel droplets in rich mixture have a limited influence on the flame ignition process. The increase in MIE $Q_c$ (marked in Fig. 2) for rich spray flames may be caused by the evaporative heat loss from evaporating fuel droplets. However, this increase is small (0.08% only), since the droplet loading is small and the heat loss from droplet evaporation is low compared to the ignition energy. The effects of droplet properties (e.g., mass loading and droplet evaporation rate) on the MIE will be further discussed later in Fig. 4. Furthermore, the fuel droplets also lead to a lower propagation speed for spray flames for a fixed flame radius (e.g., $R_f = 10$ in Figs. 2a and 2b). These tendencies are consistent with the previous work [16,17] about the effects of fuel sprays on the propagation speeds of planar ($R_f \rightarrow +\infty$) and weakly stretched spherical flames ($R_f \gg 1$).

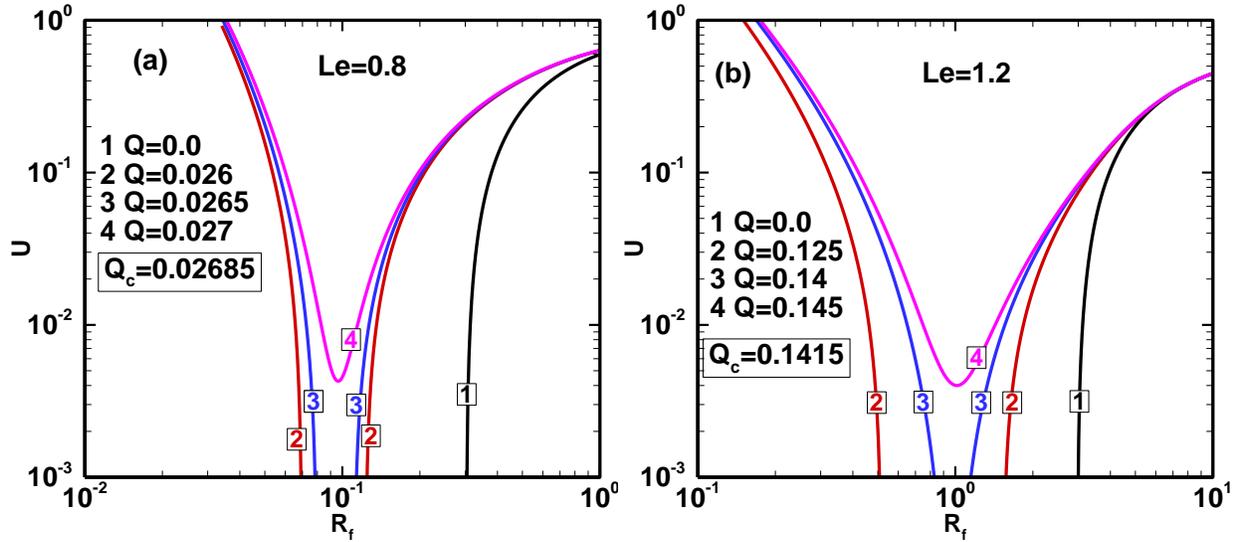

Fig. 3 Flame propagation speed as a function of flame radius with different ignition energies for (a) $Le = 0.8$ and (b) $Le = 1.2$. $\delta = 0.1$ and $\Omega = 0.2$.

The effects of Lewis number on premixed spray flame ignition are shown in Fig. 3. Two Lewis numbers are considered, i.e., 0.8 and 1.2. The droplet mass loading is $\delta = 0.1$, whereas the heat exchange coefficient is $\Omega = 0.2$, same as those with $Le = 1.0$ in Fig. 2(a). Comparing the results in Figs. 2(a) and 3, one can find that the Lewis number effects on the flame ignition process are pronounced. The Lewis number less (greater) than unity indicates the heat diffusion from the positively stretched flame front is weaker (stronger) than the mass diffusion towards it, such that the flame reactivity is enhanced (reduced) with respect to the flame with $Le = 1$ [47,48]. Therefore, the MIE when $Le = 0.8$ in Fig. 3(a) and with $Le = 1.2$ in Fig. 3(b) are respectively lower and higher than that with $Le = 1.0$ in Fig. 2(a). However, the critical ignition condition, i.e., coincidence of the two flame ball solutions ($R_Z^+ = R_Z^-$), is not affected by the Lewis number, although the critical ignition $R_{ic}$ is increased with the Lewis number.

The effects of Lewis number of the pre-vaporized gas mixtures and droplet properties on the MIE $Q_c$ are further examined in Fig. 4. It is seen that for $Le = 0.8 - 1.5$, $Q_c$ increases monotonically with $Le$, which is similar to the observations by Han and Chen [15]. This is reasonable since larger Lewis number reduce the flame reactivity and hence higher ignition energy is needed to successfully ignite the flame [32,33,44,45]. Moreover, the variations of the MIE with respect to the droplet mass loading $\delta$ and heat exchange coefficient $\Omega$ are relatively limited. As mentioned above, this is because the evaporative heat loss is relatively low at the ignition stage compared to the ignition energy deposition, due to dilute fuel droplet addition and short ignition timescale. Therefore, for successful ignition of partially pre-vaporized sprays, the critical energy is largely affected by gas phase transport properties, instead of the droplet loading and evaporation properties. Nevertheless, how the pre-vaporization degree influences the MIE necessitates detailed numerical simulations and/or experimental measurements.

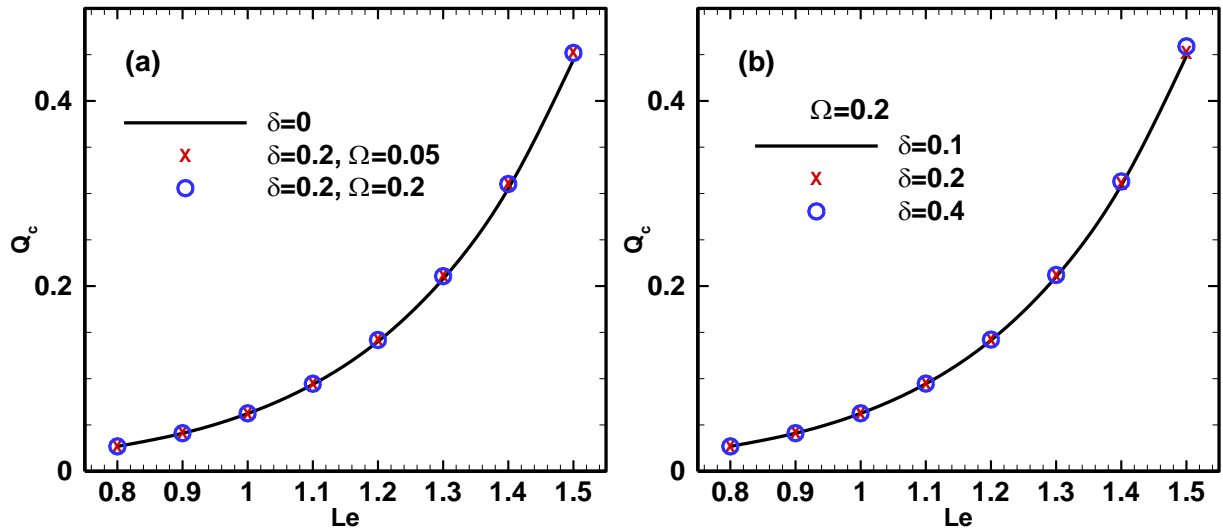

Fig. 4 Minimum ignition energy as a function of gas mixture Lewis number and droplet properties.

## 4.2 Evolution of fuel spray evaporation zone

In this section, evolutions of the droplet evaporation zone in ignition of spray flames will be investigated. To the best of our knowledge, it is the first time that fuel droplet spatial distributions are quantified and discussed in spray flame ignition through a general theory. Here we define the total radial length of the droplet evaporation zone based on the difference between the droplet onset and completion locations, i.e., $\Delta\eta = \eta_v - \eta_c$. In particular, for heterogeneous flames, the evaporation zone length in unburned and burned zones correspond to the magnitudes of $\eta_v$ and $\eta_c$, respectively.

The changes of the evaporation onset and completion locations and the evaporation zone length in spray flame ignition process are presented in Fig. 5. Here $\delta = 0.1$, $\Omega = 0.2$, and $Le = 1.0$. The corresponding flame propagating speeds have been shown in Fig. 3(a). As clarified in section 2.3, what our theory can describe starts from the instant when the droplet evaporation completion front $\eta_c$ critically propagate off the ignition location, which corresponds to the dashed line of $\eta_c = -R_f$ as in Fig. 5(b). When the ignition energy $Q$ is less than the MIE $Q_c$ (e.g., lines #2 −#3), the droplet onset and completion locations ($\eta_v$ and $\eta_c$) have left and right branches (see Figs. 5a and 5b). They respectively correspond to the flame kernel and propagating branches. One can see that droplet evaporation onset location $\eta_v$ is positive, indicating that the droplets start to vaporize in the fresh gas when they are heated to boiling temperature. It monotonically increases with the flame radius along the flame kernel branch, which means that the distance between the flame front ($\eta = 0$) and evaporation onset location increases. This is justifiable because the reactivity of the flame kernel is gradually reduced as the effects of the ignition energy fade, and accordingly the temperature gradient in the pre-heat zone is decreased, resulting in a farther location where the boiling temperature can be critically reached [17].

For the evaporation completion front $\eta_c$ corresponding to the flame kernels, when the ignition energy is added, the droplets are distributed in the whole domain (parameterized by $\eta_c = -R_f$), as demonstrated in Fig. 5(b). As the evaporation of the fuel droplets near the spark proceeds caused by the locally high temperature, the evaporation completion front moves outwardly, following the leading flame front. This corresponds to the condition of $-R_f < \eta_c < 0$, and the mixture around the flame front is heterogeneous (vapour and liquid droplets). This can be clearly seen in the inset of Fig. 5(b). However, like along lines #2 and #3, as the flame propagation speed rapidly decreases and the droplet evaporation is accelerated, the evaporation completion front catches up with and crosses the flame front, and ultimately propagates before the flame front. Hence, along the flame kernel branch, the mixture around the reactive front experiences a transition from heterogeneous to homogeneous condition. One can see that this transition corresponds to the zero-crossing points (intersection points of the $\eta_c$ curve and flame front FF in the inset of Fig. 5b). Afterwards, the flame kernel grows in a purely gaseous mixture, until it is degraded to a stationary flame ball (black dots in Fig. 5b).

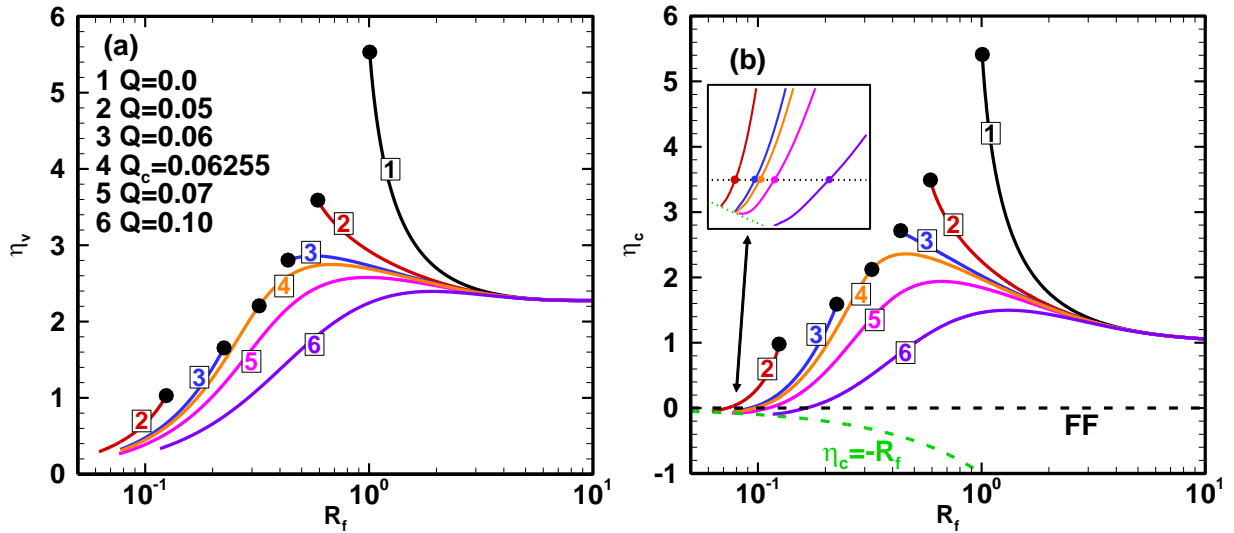

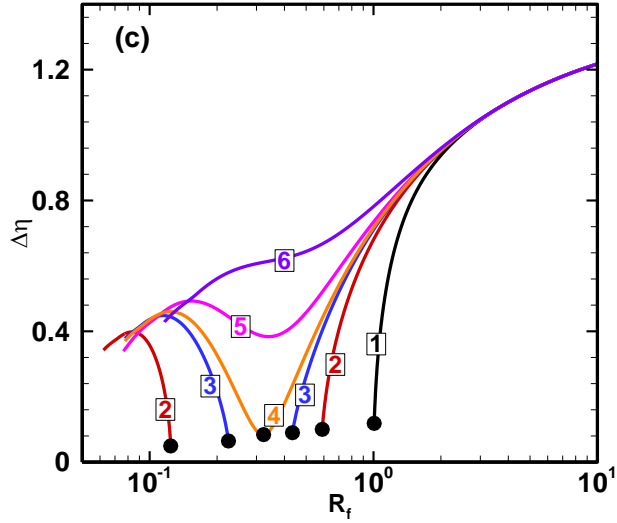

Fig. 5 (a) Evaporation onset location, (b) evaporation completion location, and (c) evaporation zone length as functions of flame radius. $\delta = 0.1$, $\Omega = 0.2$, $Le = 1.0$. Black dot: flame ball. FF: flame front. Colored points in the inset of Fig. 5(b): flame transition points.

For the right flame propagation branch, the reactant mixtures near the flame front are homogeneous, since the droplets have been vaporized before the flame front (i.e., $\eta_c > 0$). Both evaporation onset and completion locations decrease as the homogeneous flames expand, as shown in Figs. 5(a) and 5(b). This is due to the gradual enhancement of the spherical spray flame caused by the promoted convective−diffusion process when it propagates outwardly [33], manifested by the increased propagation speed in Fig. 2(a). As such, the droplets are heated to boiling temperature closer to the flame front, due to the large temperature gradient in the fresh gas mixture [17].

Figure 5(c) shows the radial length of the evaporation zone as a function of the flame radius. When the ignition energy is lower than the MIE (lines #2 and #3), for the left flame kernel branch, the evaporation zone length $\Delta\eta$ first increases and quickly decreases in the flame kernel decaying process. However, along the flame propagation branch, the evaporation zone length monotonically increases with the flame radius. When the ignition energy is larger than MIE (lines #4−#6), the evaporation zone length becomes non-monotonic near the critical ignition radius $R_{ic}$ (about 0.3), but this nonmonotonicity is gradually weakened as the ignition energy increases (e.g., line #6). The length of the evaporation zone is expected to have direct influence on the heat absorption from the gas phase, which will be further discussed in Section 4.4.

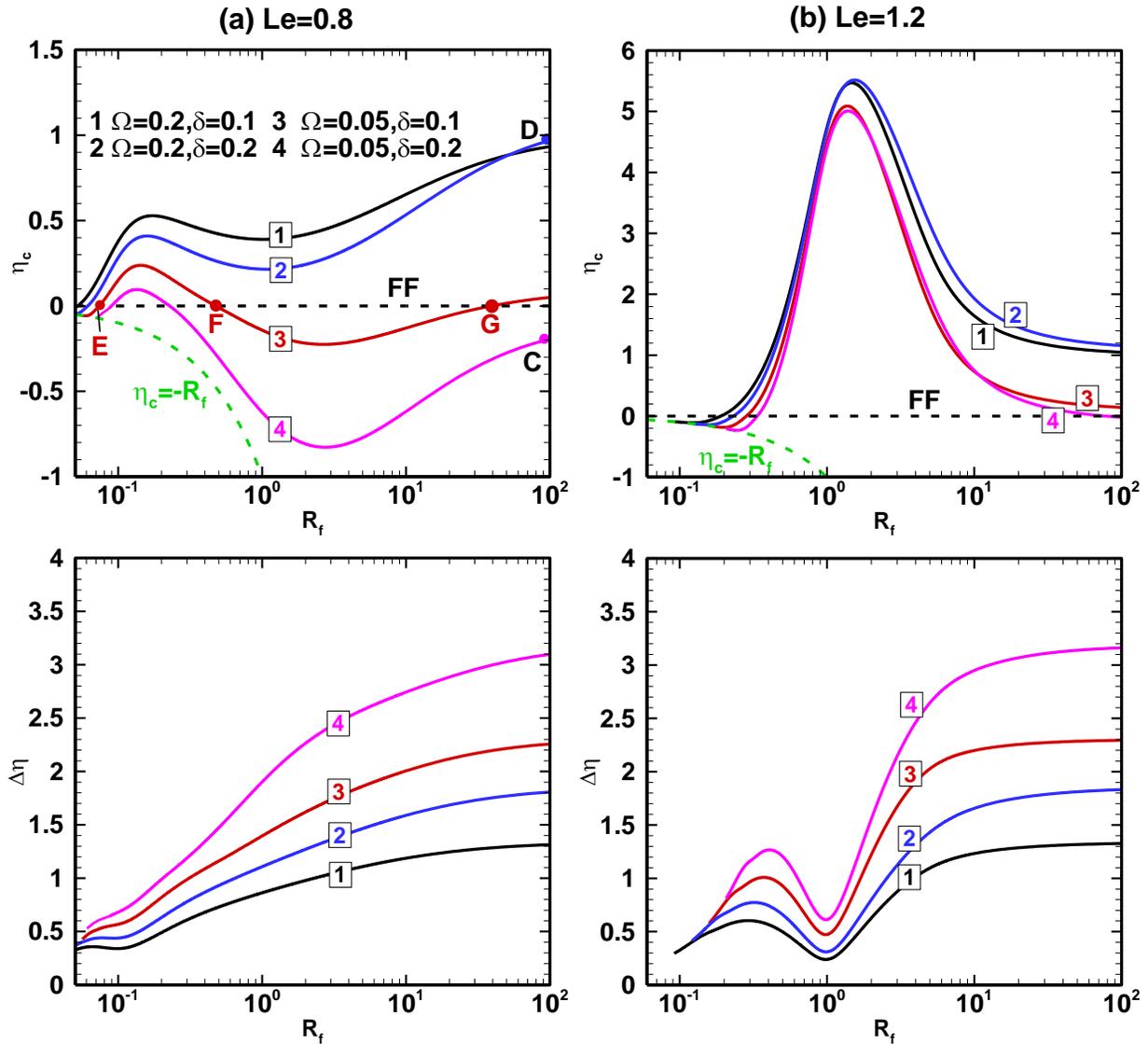

Fig.6 Evaporation completion location and evaporation zone length as functions of the flame radius with different $\delta$ and $\Omega$: (a) $Le$ =0.8 and (b) $Le$ = 1.2. FF: flame front.

The variations of evaporation completion location $\eta_c$ and evaporation zone length $\Delta\eta$ with flame radius $R_f$ under different Lewis number and droplet properties are shown in Figs. 6(a) and 6(b). Ignition energies above the respective MIE are used to successfully ignite spray flames under these two conditions (i.e., $Q = 0.03$ and $0.145$ for Figs. 6a and 6b, respectively). For $Le = 0.8$ in Fig. 6(a), heat diffusion from the spherical flame front is weaker compared to species diffusion towards it. Therefore, the spherical flame is stronger and hence quicker to transit from highly stretched igniting kernel to stably propagating flame. The evaporation completion location $\eta_c$ versus flame radius in Fig. 6(a) is *N*-shaped: firstly, increases until the critical ignition radius $R_{ic}$, then decreases, and further increases to the values of propagating flames. Multiple transitions between HT flame to HM flame occur when the heat exchange coefficient (hence droplet evaporation rate) is low (e.g., $\Omega = 0.05$ in lines #3 and #4), characterized by multiple flame transition (zero-crossing) points. For example, along line #3, the mixture near the kernel is initially heterogeneous ($\eta_c < 0$, left to point E, droplets + vapor), and becomes homogeneous ($\eta_c > 0$, vapor only) between points E and F when the flame propagates outwardly. For subsequent propagation after point F, it is heterogeneous (Curve FG), but after point G, homogeneous combustion with complete gasification of droplets in the preheat area is observed again. For increased droplet loading $\delta$ or decreased heat transfer $\Omega$ (hence weaker evaporation rate $\omega_v$, see Eq. 13), it is more difficult for the gaseous flame to fully vaporized the droplets in the pre-flame zone. Therefore, increased $\delta$ and/or decreased $\Omega$ (also $\omega_v$) would make the spray flame tend to be HT as shown in Fig. 6(a). Besides, the evaporation zone length $\Delta\eta$ almost monotonically increases, regardless of the foregoing transitions between homogeneous and heterogeneous flames.

For $Le = 1.2$ in Fig. 6(b), the evaporation completion location curve is approximately $\Lambda$-shaped. Initially, the flame is heterogeneous, and the droplet completion front moves closer to and crosses the flame front, leading to a homogeneous flame with varying evaporation completion location $\eta_c$. Transitions between HM and HT combustion are also seen at low heat exchange coefficient and evaporation rates (e.g., line #4), similar to what is found in Fig. 6(a). Near the critical ignition radius $R_{ic}$ (around 1), fuel droplets can start to vaporize farthest in the preheat area relative to the flame front, but the evaporation zone length is small, regardless of the droplet loading and evaporation rate. This is due to the weakened reactivity of the flame, manifested by the low propagating speed, as shown in Fig. 3(b). Meanwhile, the evaporation completion front $\eta_c$ is much higher than that in Fig. 6(a) with $Le$ = 0.8. This is due to the enhance heat diffusion from the flame front due to high Lewis number, which leads to more distributed gas temperature in the two-phase mixture ahead of the flame and farther fresh gas with boiling temperature. Based on the results in both Figs. 6(a) and 6(b), one can also see that the evaporation zone length $\Delta\eta$ generally increases with droplet loading, whilst decreases with heat exchange coefficient (or droplet evaporation rate).

Figures 6(a) and 6(b) also indicate that for gaseous mixtures with large Lewis number and/or rapidly evaporating liquid fuels, the flame front generally propagates in a gaseous (e.g., fuel vapor and oxidizer) environment. Understanding this feature through our analysis would be of great importance to accurately and effectively model the interactions between the chemistry and droplet evaporation, particularly for turbulent spray combustion. Whether the droplet evaporation effects are needed to be incorporated in advanced combustion models is still an open question. For instance, Mortensen and Bilger derived the Conditional Moment Closure (CMC) combustion model for spray flame, through considering the droplet evaporation in the mixture fraction space [49]. This leads to more sophisticated implementations and modelling issues, see Refs. [50,51], compared to the CMC model for gaseous flame [52]. In reality, before their work, the latter has also been successfully used for modelling spray combustion with reasonable accuracies [53]. Similar problems also exist in other combustion modelling approach, e.g., tabulated chemistry method [54]. Therefore, identifications of homogeneous and heterogeneous flames in spray combustion through theoretical analysis can provide the general evidence for selecting physically sound combustion model for a range of problems.

### 4.3 Stationary flame ball in liquid fuel sprays

Due to the important role of flame balls in determining the critical conditions for successful flame ignition as unveiled in Figs. 2 and 3, the droplet evaporation characteristics of the flame balls are worthy of further discussion. In Fig. 5, the flame ball solutions are marked with black dots. We can find that for different ignition energies the inner and outer flame balls always exist in homogeneous mixtures. It is known that the diffusion-controlled flame balls are sustained by the transport of heat and species with the ambient gas [55]. In two-phase mixture, the post-flame droplet evaporation is unsteady, and hence it is difficult to maintain steady heat and mass transfer for stationary flame balls.

The droplet evaporation completion location $\eta_c$ and evaporation zone length $\Delta\eta$ corresponding to different stationary flame ball radii $R_z$ in Fig. 5 are shown in Fig. 7. The Lewis number is fixed to be $Le = 1.0$. The reader is reminded that for different ignition energies $Q$ below the MIE, there are inner and outer flame ball solutions ($R_Z^+$ and $R_Z^-$), as mention in Fig. 2. In Fig. 7(a), the curves of evaporation completion location and evaporation zone length can be divided into two sections by the MIE points: the left denotes the solutions for inner flame balls $R_Z^+$ from the igniting kernel branch, while the right one $R_Z^-$ from the propagating branch. With increased ignition energy $Q$ (marked with "$Q \uparrow$" in Fig. 7a) from the two ends of the curves (marked with $Q = 0$ and $Q \approx 0$ in Fig. 7a), $R_Z^+$ and $R_Z^-$ respectively move towards each other.

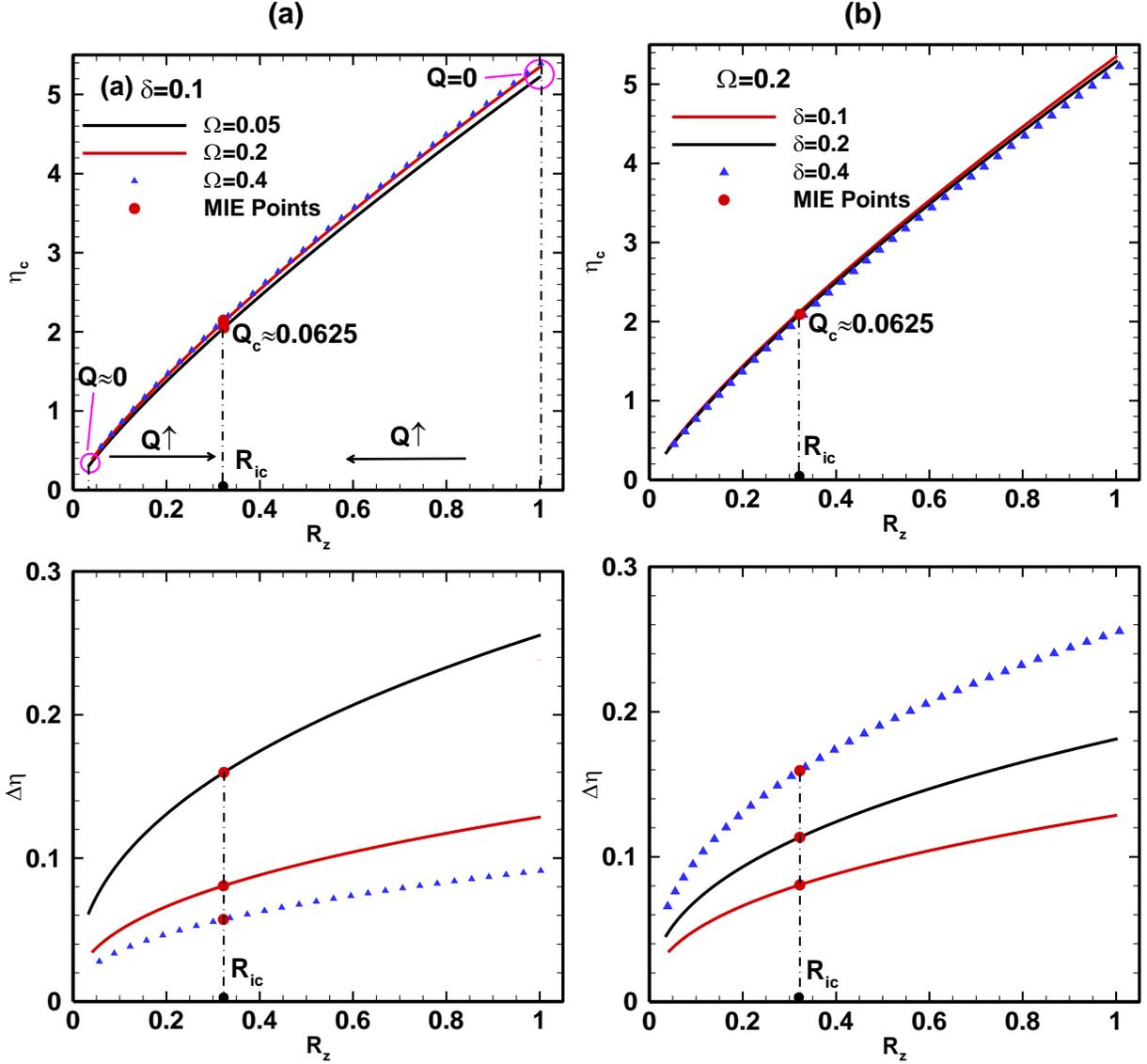

Fig. 7. Evaporation completion location and evaporation zone length for the flame ball under different ignition energies for (a) $\delta = 0.1$ and (b) $\Omega = 0.2$. $Le = 1.0$. $R_{ic}$: critical ignition radius.

One can see from Fig. 7 that for the inner (outer) flame balls, $\eta_c$ monotonically increases (decreases). This is also true for the evaporation zone length $\Delta\eta$. The critical ignition condition (i.e., $R_Z^+ = R_Z^-$) is reached at intermediate values of $\eta_c$ and $\Delta\eta$. For the inner flame balls located at the flame kernel branch (left to MIE points), the flame temperature is affected by ignition energy. For larger ignition energy, the flame temperature and gas temperature in the pre-flame zone are higher. Hence, the spray droplets can finish its evaporation farther away from the flame front, which indicates the $\eta_c$ increases with the ignition energy for the inner flame balls. Meanwhile, as seen in Fig. 2(a), the flame ball radius moves outwardly with increased ignition energy. The evaporation zone length also increases due to simultaneous increase of the evaporation onset location. For the outer flame balls in the flame propagating branch (right to MIE points), the gas temperature in the pre-flame zone decreases with ignition energy, resulting in decreased flame ball radius as shown in Fig. 2(a). Therefore, opposite trends are observed for them.

For fixed droplet loading $\delta$ = 0.1 in Fig. 7(a), increase in heat exchange coefficient (hence evaporation rate) results in limited variations in evaporation completion fronts for the same flame ball radius. However, the evaporation zone length $\Delta\eta$ of the stationary flame balls has considerably decreased with increased $\Omega$, because of larger heat exchange coefficient and faster evaporation rate. As for fixed heat exchange coefficients $\Omega$ = 0.2 in Fig. 7(b), it is found that increasing droplet loading would induce the longer evaporation zone since the total amount of fuel droplet is increased while the evaporation rate remains constant for the same heat exchange coefficient. The critical ignition radius, $R_{ic}$, almost remains unchanged with varied droplet properties for fixed Lewis number.

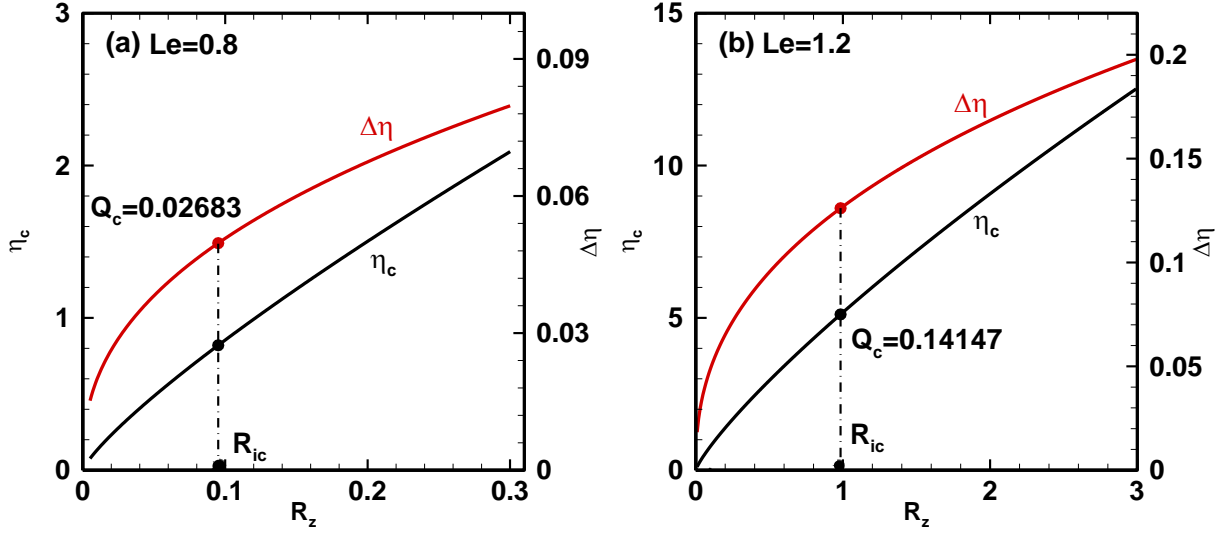

Fig. 8. Evaporation completion location and evaporation zone length for flame balls under different ignition energies for $\delta = 0.1$, $\Omega = 0.2$ when (a) $Le = 0.8$ and (b) $Le = 1.2$.

The Lewis number effects on the droplet evaporation characteristics of the flame balls are examined in Fig. 8. For varied Lewis number with fixed droplet properties, i.e., $\delta = 0.1$ and $\Omega = 0.2$, as shown in Fig. 7 and Fig. 8, the critical ignition radius $R_{ic}$ increases from 0.095 to 0.984 as $Le$ increased from 0.8 to 1.2. Meanwhile, the evaporation completion front $\eta_c$ and the evaporation zone length $\Delta\eta$ of the flame balls also increases as $Le$ increases.

As shown in Figs. 7 and 8, critical successful ignition always occurs through flame balls in homogeneous mixtures ($\eta_c > 0$). Therefore, the relations between the timescales of droplet evaporation, flame propagation and spark ignition become important. For instance, if more slowly evaporating droplets are dispersed near the spark, longer spark addition is required to achieve the homogeneous mixture for flame kernel formation and growth. Moreover, the transition from initial heterogeneous flame to homogeneous one necessitates faster propagation of the-phase contact surface compared to the flame front. There have been some studies about these relations based on experimental measurements [5] and direct numerical simulations [10]. However, how these timescales determine the critical ignition merits further investigations.

## 4.4 Flame structure and evaporation heat loss

In Figs. 9(a) and 9(b), the flame structures for HT ($R_f = 0.0776$) and HM ($R_f = 0.2249$) flame along the flame kernel branch are presented, which are also marked in Fig. 2(a). The reader is reminded that Fig. 9(b) corresponds to a stationary flame ball. Here the droplet and gas phase parameters are $\delta = 0.1$, $\Omega = 0.2$, $Q = 0.06$ and $Le = 1.0$. For both, the effects of the ignition energy are still significant, which is manifested by the high temperature close to the spherical centre ($r = 0$). Meanwhile, the oxidizer is fully consumed at the flame front $R_f$, and hence there is no oxidizer in the burned zone ($r < R_f$). For the HT flame in Fig. 9(a), the fuel droplets penetrate through the flame front into the burned zone, corresponding to the finite values of local droplet loadings $Y_d$. However, for the HM flame in Fig. 9(b), the evaporation completion front $R_c$ is located before the flame front $R_f$ in Fig. 9(b) and therefore two-phase mixture only exists before the flame front. These lead to different influences on heat loss of the gas phase due to droplet evaporation, which will be further quantified in Figs. 11 and 12.

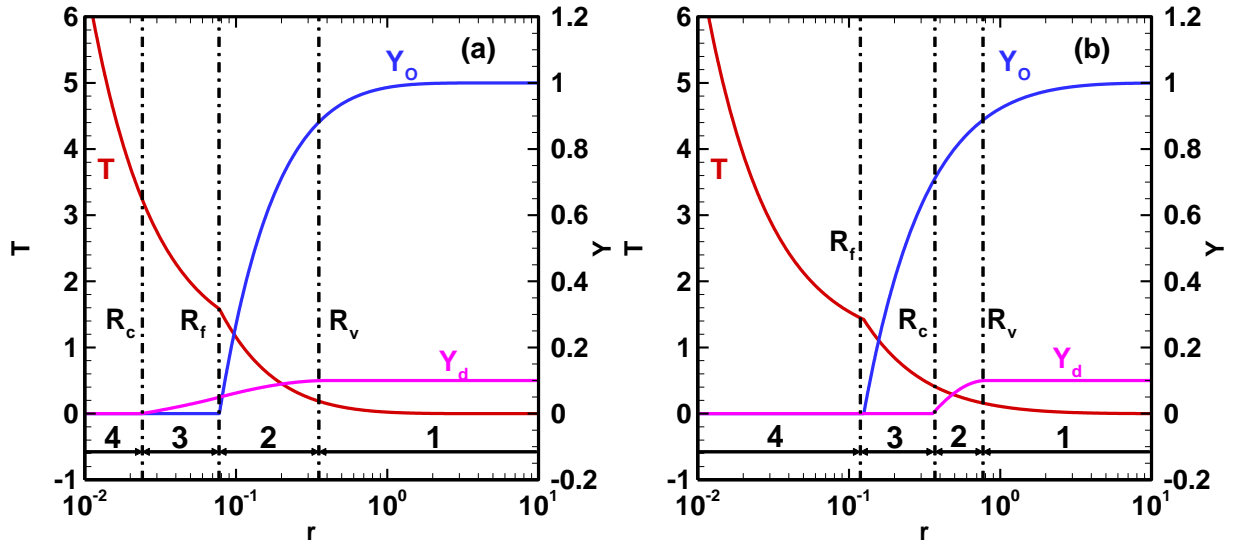

Fig. 9. Spatial distributions of temperature, oxidizer mass fraction, and droplet mass loading with flame radii $R_f =$ (a) 0.0776 (point A in Fig. 2a) and (b) 0.2249 (point B in Fig. 2a). $\delta = 0.1$, $\Omega = 0.2$, $Q = 0.06$, and $Le = 1.0$. $R_{v,c} = \eta_{v,c} + R_f$.

In Figs. 10(a) and 10(b), the structures for HT ($\Omega = 0.05$) and HM ($\Omega = 0.2$) flames from the propagation branch are presented. The droplet and gas phase properties are $\delta = 0.1$, $R_f = 100$, $Q = 0.06$ and $Le = 0.8$, respectively. They are marked as C and D in Fig. 6(a). The gas temperature in the burned zone is uniform and below unity due to the evaporation heat absorption. Meanwhile, the flame initiation effect from the ignition energy disappears. For HT and HM flames, their evaporation completion front $R_c$ lies behind and before the flame front $R_f$, respectively.

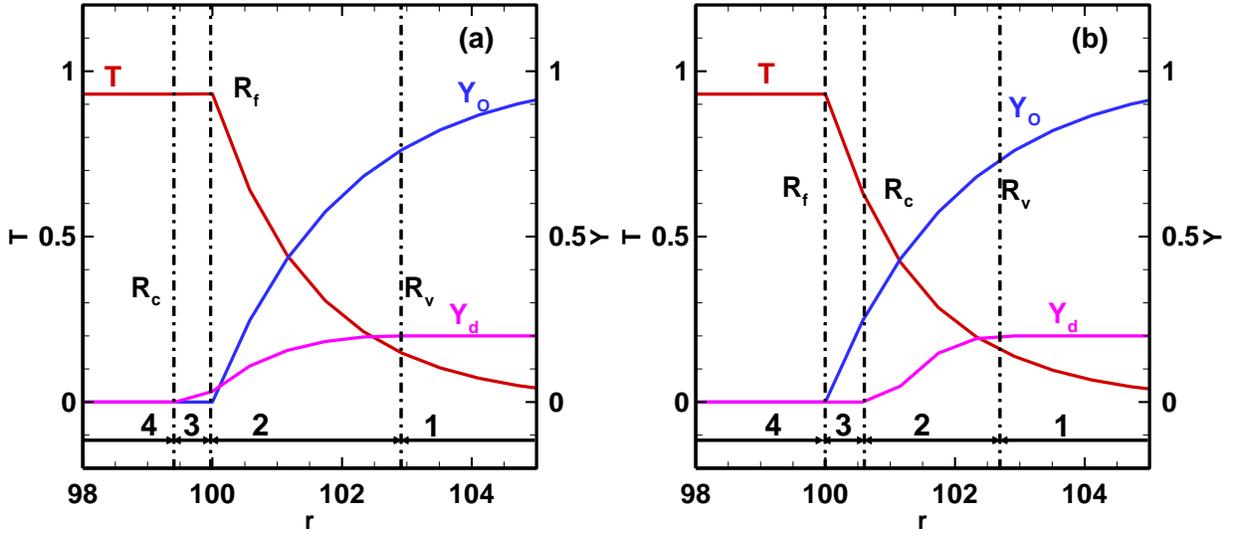

Fig. 10. Spatial distributions of temperature, oxidizer mass fraction and droplet mass loading with heat exchange coefficients $\Omega$ = (a) 0.05 (point C in Fig. 6a) and (b) 0.2 (point D in Fig. 6a). $\delta = 0.2$, $R_f = 100$, $Q = 0.03$, and $Le = 0.8$.

As discussed above, HT and HM flames can exist in both flame kernel and flame propagation stages. Therefore, it would be interesting to quantify the effects of droplet heat absorption from the gas phase corresponding to various flame regimes and development stages. We define the normalized evaporative heat loss based on the gas temperature profiles in the burned and unburned zones of a HT flame as [17,21]

$$H_{ub,HT} = \Omega \int_0^{\eta_v} [T_2(\xi) - T_v](\xi + R_f)^2 d\xi \bigg/ \left( R_f^2 \frac{dT}{d\eta} \bigg|_- - R_f^2 \frac{dT}{d\eta} \bigg|_+ \right), \qquad (42)$$

$$H_{b,HT} = \Omega \int_{\eta_c}^{0}[T_3(\xi) - T_v](\xi + R_f)^2 d\xi / \left(R_f^2 \frac{dT}{d\eta}\Big|_{-} - R_f^2 \frac{dT}{d\eta}\Big|_{+}\right). \tag{43}$$

Likewise, for a HM flame, they respectively are

$$H_{ub,HM} = \Omega \int_{\eta_c}^{\eta_v}[T_2(\xi) - T_v](\xi + R_f)^2 d\xi / \left(R_f^2 \frac{dT}{d\eta}\Big|_{-} - R_f^2 \frac{dT}{d\eta}\Big|_{+}\right), \tag{44}$$

$$H_{b,HM} = 0. \tag{45}$$

The subscripts "*ub*" and "*b*" respectively denote evaporative heat loss from unburned and burned zones, whereas "HM" and "HT" respectively denote homogeneous and heterogeneous flames. The denominator, $\left(R_f^2 \frac{dT}{d\eta}\Big|_{-} - R_f^2 \frac{dT}{d\eta}\Big|_{+}\right)$, is the combustion heat release. Equation (45) is valid since there are no droplets in the burned zone of homogeneous flames. Accordingly, the total evaporation heat loss of a spray flame is $H_{all} = H_b + H_{ub}$.

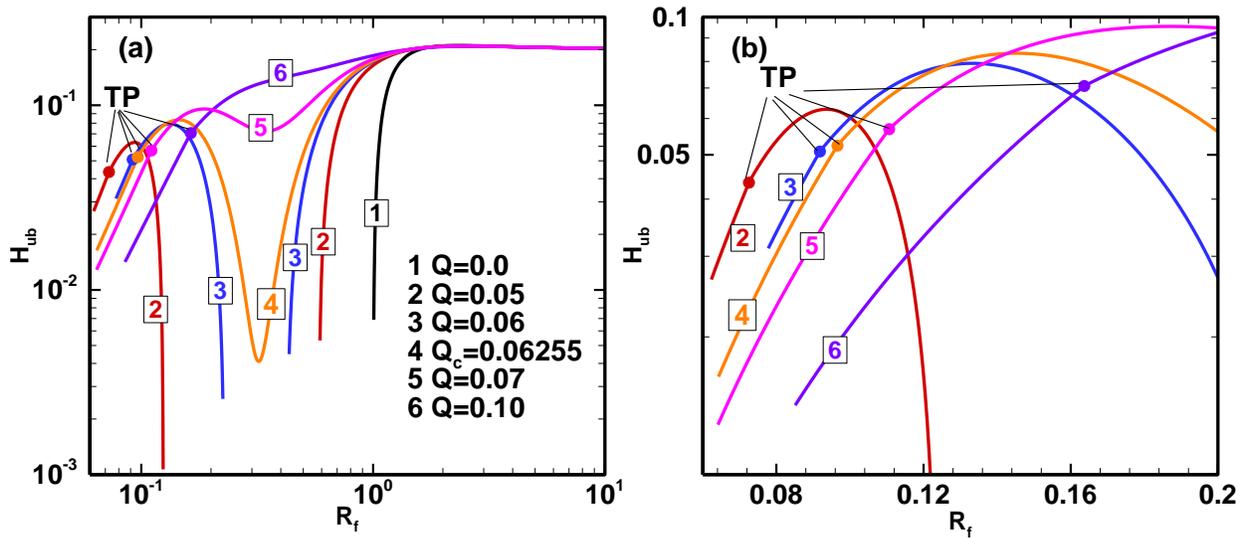

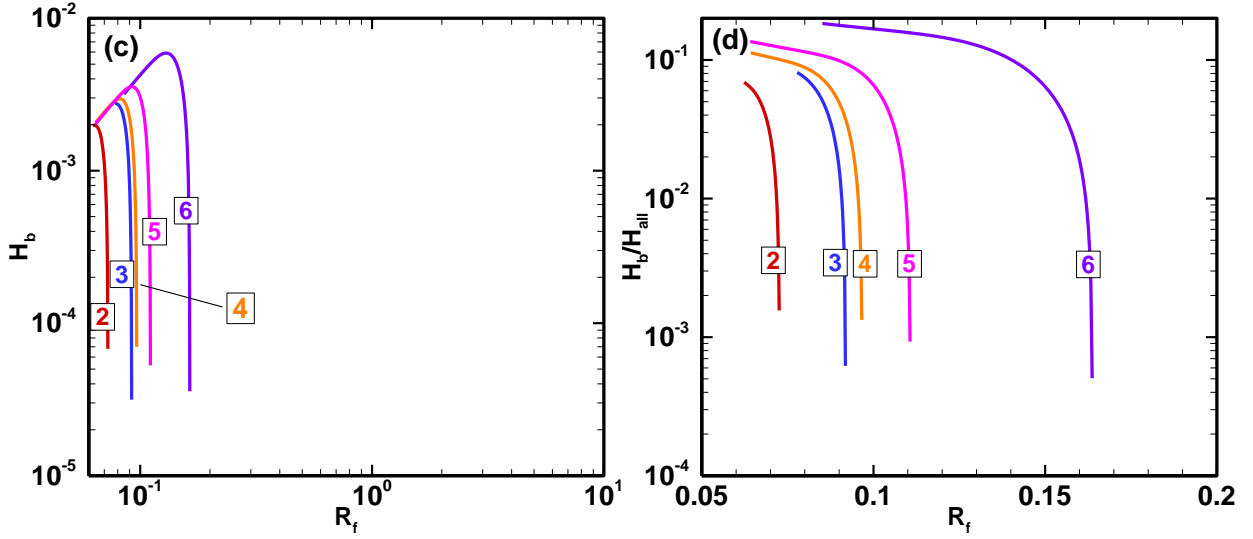

Fig. 11. Evaporative heat loss in the unburned and burned zones as a function of flame radius $R_f$ with different ignition energies. $\delta = 0.1$, $\Omega = 0.2$ for $Le = 1.0$. TP: Flame regime transition points.

Figures 11(a)−11(c) show that the evaporative heat loss for spray flames with different ignition energies in Fig. 2(a) when $\delta = 0.1$, $\Omega = 0.2$, and $Le = 1.0$. The enlarged profiles of $H_{ub}$ around the TP points are plotted in Fig. 11(b). One can see from Fig. 11(a) that, when the ignition energy is less than the MIE, the evaporative heat loss of the flame kernel first increases and then decreases, as shown in lines #2 and #3. This is associated with the gradually reduced ignition energy dissipation to vaporize the surrounding fuel droplets, when the ignition kernel grows. The reader is reminded that the mixture changes from heterogeneous ($R_f < R_{TP}$) to homogeneous ($R_f > R_{TP}$) conditions at the transition point TP. Continuous variations of $H_{ub}$ across the TP's can be seen in Figs. 11(a) and 11(b). For the flame propagation branch (lines #2 and #3), $H_{ub}$ increases when the flame propagates outwardly. When the flame is successfully ignited, $H_{ub}$ generally increases, although a reduction can be seen near the critical ignition radii $R_{ic}$. The evaporative heat loss in the post-flame area of heterogeneous flames (before TP) are shown in Fig. 11(c). One can see that $H_b$ first increases and then decreases when the flame is close to the TP points. The increases are caused by the enhanced droplet evaporation due to the spark, whilst the decrease results from the reduced evaporation zone length $\Delta\eta$. It should be noted that $H_b$ at the transition points should be zero, which is not shown due to the logarithmic scale used.

Plotted in Fig. 11(d) are the ratios of the evaporation heat loss in the burned zone $H_{b,HT}$ to the total one $H_{all}$ in HT flames. The ratio decreases from approximately 0.1 to 0.001 when the flame kernel propagates. This is because the post-flame evaporation zone gradually diminishes as the evaporation completion front moves faster than the leading flame front. Besides, the geometry effect of $(\xi + R_f)^2$ in Eqs. (42) – (44) becomes prevalent when the flame radius increases. In general, the post-flame evaporative heat loss has small contributions when the flame is initially developed. In this sense, the evaporation heat loss in the unburned zone is dominant, although it is essentially affected the evaporation zone length.

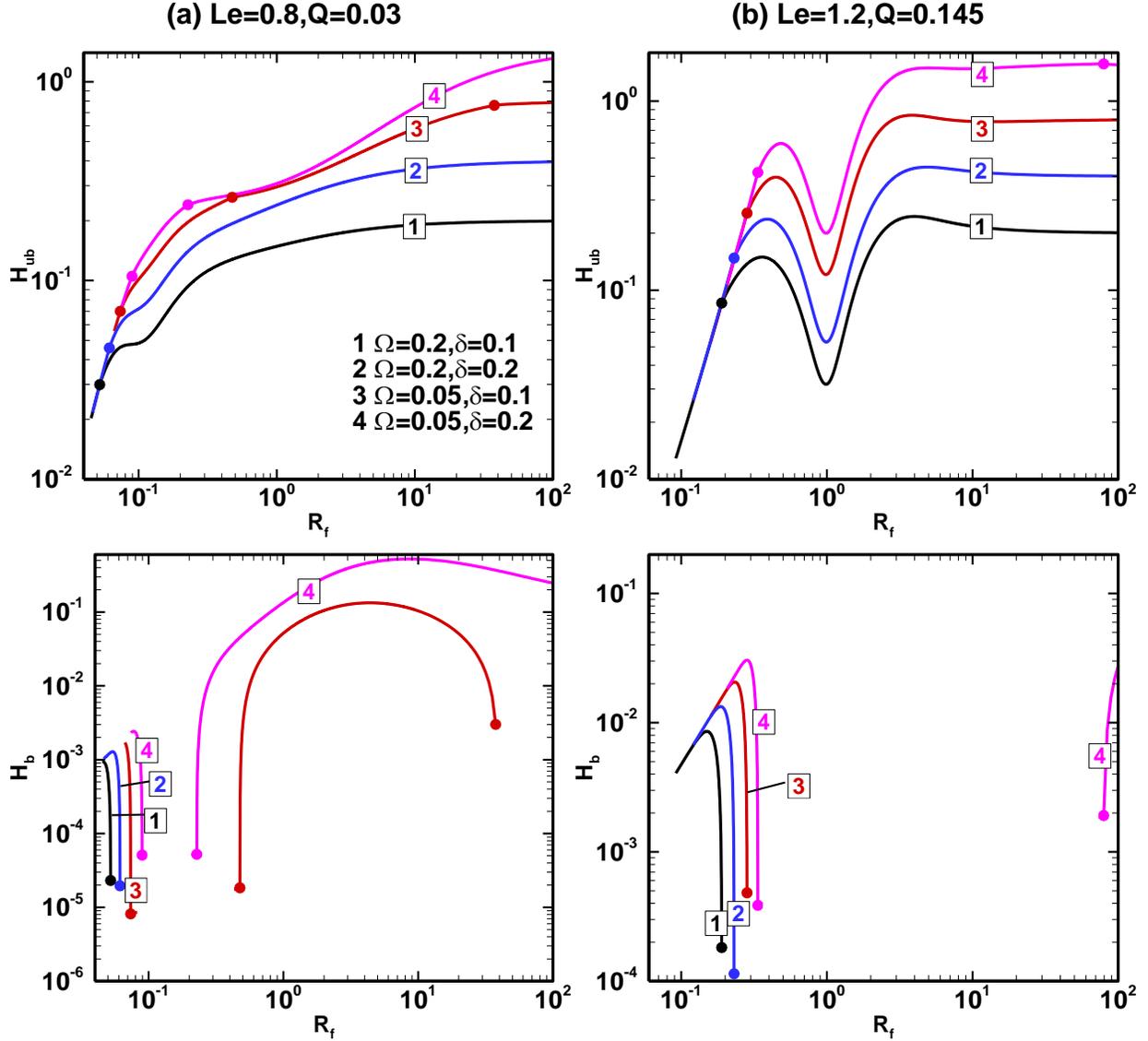

Fig. 12. Evaporative heat loss in the unburned and burned zone as a function of flame radius $R_f$ with different $\delta$ and $\Omega$ for (a) $Le$ = 0.8, $Q$ = 0.03, and (b) $Le$ = 1.2, $Q$ = 0.145. Colored points: TP.

The effects of Lewis number and droplet properties on the evaporative heat loss are presented in Fig. 12. The evolutions of the flame propagating speed have been shown in Fig. 6. One can find that the evaporation heat loss in the pre-flame zones $H_{ub}$ with respect to the flame radius is correlated to the evaporation zone length $\Delta\eta$, which is shown in Fig. 8. Meanwhile, the evaporative heat loss in both pre- and post-flame zones is increased with increased droplet loading or decreased heat exchange coefficient. This is because increased droplet loading indicates that more heat has been transferred into the liquid phase for evaporation, while decreased heat exchange coefficient means that the evaporation zone length is extended due to small evaporation speed. Additionally, the change of the heat loss in the burned zone of heterogeneous flame kernels is similar to that in Fig. 11(c). However, heterogeneous flames can also be observed when the flame radius is large, like lines #3 and #4 in Fig. 12(a) and line #4 in Fig. 12(b). Their respective heat loss, $H_b$, is comparable to $H_{ub}$ at the same flame radius. The trend of $H_b$ generally follows that of the magnitude of the evaporation completion location $\eta_c$ for stably propagating HT flames (see Fig. 6).

# 5 Conclusion

Evolution of droplet evaporation zone and its interaction with the propagating flame front are studied in this work. A general theoretical model is derived to describe ignition and propagation of fuel-rich premixed spray flames in homogenous and heterogeneous mixtures. The change of droplet evaporating zones with reaction front and transitions between homogeneous and heterogeneous flames are considered. The correlations between flame propagating speed, flame temperature, and characteristic droplet evaporation locations are numerically calculated to elucidate the effects of various gas and liquid properties.

The results show that the spray flame trajectories are considerably affected by ignition energy. Moreover, the critical condition for successful ignition is coincidence of inner and outer flame balls from flame kernel and propagating branches. Moreover, the minimum ignition energy is negligibly affected by droplet mass loading and heat exchange coefficient under fuel-rich mixture conditions. However, it increases monotonically with the Lewis number.

It is also found that the flame kernel always originates from heterogeneous mixtures due to the initially dispersed droplets near the spark. However, transitions from heterogeneous and homogeneous mixtures occurs due to rapidly droplet gasification. Under some conditions, multiple transitions between the two flame regimes can be seen in the stages of igniting kernel growth and flame expansion. Before successful ignition, droplet evaporation onset and completion locations increase and decrease respectively with the flame radius. For successful flame ignition, non-monotonic variations of the above locations are observed. We also see that the flame balls always exist in homogeneous mixtures, indicating that failed and critical successful ignition events occur only in purely gaseous mixture near the flame front. The evaporating droplet distributions for flame balls are also discussed and it is shown that when the ignition energy is increased towards the minimum ignition energy, the evaporation zone length for inner and outer flame balls increases and decreases, respectively.

The evaporative heat loss of homogeneous and heterogeneous spray flames is calculated, and the results show that for the failed flame kernels, heat loss from droplet evaporation behind and before the flame front first increases and then decreases, due to the gradually fading ignition energy effects. The evaporative heat loss before the flame front generally increases, although non-monotonicity exists, when the flame is successfully ignited and propagate outwardly. In particular, for heterogeneous flames, the ratio of the heat loss from burned zone to the total one decreases as the flame expands. Moreover, droplet mass loading and heat exchange coefficient considerably affect the evaporating heat loss from both burned and unburned zones.

When our theoretical model is derived, assumptions and simplifications are used for gas and droplets alike, such as constant thermal properties and interphase thermal/kinematic equilibria. Therefore, the conclusions are only applicable for spray flames approximately satisfying the above conditions. To achieve more quantitatively accurate descriptions of gas and liquid phases, detailed numerical simulations are necessitated, which will be interesting topic for our future studies.

**Declaration of Competing Interest**

This research did not receive any specific grant from funding agencies in the public, commercial, or not-for-profit sectors.

**Acknowledgement**

QL is supported by NUS Research Scholarship. The calculations are performed with the ASPIRE 1 Cluster from National Supercomputing Center in Singapore (https://www.nscc.sg/).

**References**

[1] E. Mastorakos, Forced ignition of turbulent spray flames, Proc. Combust. Inst. (2017).
[2] S.K. Aggarwal, A review of spray ignition phenomena: present status and future research, Prog. Energy Combust. Sci. 24 (1998) 565–600.
[3] E. Mastorakos, Ignition of turbulent non-premixed flames, Prog. Energy Combust. Sci. 35 (2009) 57–97.
[4] T. Marchione, S.F. Ahmed, E. Mastorakos, Ignition of turbulent swirling n-heptane spray flames using single and multiple sparks, Combust. Flame. 156 (2009) 166–180.
[5] P.M. de Oliveira, P.M. Allison, E. Mastorakos, Ignition of uniform droplet-laden weakly turbulent flows following a laser spark, Combust. Flame. 199 (2019) 387–400.
[6] F. Akamatsu, K. Nakabe, M. Katsuki, Y. Mizutani, T. Tabata, Structure of Spark-Ignited Spherical Flames Propagating in a Droplet Cloud, in: Dev. Laser Tech. Appl. to Fluid Mech., Springer, 1996: pp. 212–223.
[7] L. Fan, B. Tian, C.T. Chong, M.N.M. Jaafar, K. Tanno, D. McGrath, P.M. de Oliveira, B. Rogg, S. Hochgreb, The effect of fine droplets on laminar propagation speed of a strained acetone-methane flame: Experiment and simulations, Combust. Flame. 229 (2021) 111377.
[8] F. Atzler, M. Lawes, S.A. Sulaiman, R. Woolley, Effects of droplets on the flame speed of laminar Iso-octane and air aerosols, 10th Int. Conf. Liq. At. Spray Syst. ICLASS 2006. (2006).

[9]     P.M. de Oliveira, E. Mastorakos, Mechanisms of flame propagation in jet fuel sprays as revealed by OH/fuel planar laser-induced fluorescence and OH* chemiluminescence, Combust. Flame. 206 (2019) 308–321.

[10]    A.P. Wandel, N. Chakraborty, E. Mastorakos, Direct numerical simulations of turbulent flame expansion in fine sprays, Proc. Combust. Inst. 32 (2009) 2283–2290.

[11]    G. Ozel Erol, J. Hasslberger, M. Klein, N. Chakraborty, A direct numerical simulation analysis of spherically expanding turbulent flames in fuel droplet-mists for an overall equivalence ratio of unity, Phys. Fluids. 30 (2018).

[12]    A. Neophytou, E. Mastorakos, R.S. Cant, DNS of spark ignition and edge flame propagation in turbulent droplet-laden mixing layers, Combust. Flame. 157 (2010) 1071–1086.

[13]    R. Thimothée, C. Chauveau, F. Halter, I. Gökalp, Experimental investigation of the passage of fuel droplets through a spherical two-phase flame, Proc. Combust. Inst. 36 (2017) 2549–2557.

[14]    J.B. Greenberg, Finite-rate evaporation and droplet drag effects in spherical flame front propagation through a liquid fuel mist, Combust. Flame. 148 (2007) 187–197.

[15]    W. Han, Z. Chen, Effects of finite-rate droplet evaporation on the ignition and propagation of premixed spherical spray flame, Combust. Flame. 162 (2015) 2128–2139.

[16]    Q. Li, H. Zhang, C. Shu, Propagation of heterogeneous and homogeneous planar flames in fuel droplet mists, Int. J. Multiph. Flow. 133 (2020) 103452.

[17]    Q. Li, H. Zhang, C. Shu, Propagation of weakly stretched premixed spherical spray flames in localized homogeneous and heterogeneous reactants, Phys. Fluids. 32 (2020) 123302.

[18]    Y. Zhuang, H. Zhang, On flame bifurcation and multiplicity in consistently propagating spherical flame and droplet evaporation fronts, Int. J. Multiph. Flow. (2020) 103220.

[19]    N.S. Belyakov, V.I. Babushok, S.S. Minaev, Influence of water mist on propagation and suppression of laminar premixed flame, Combust. Theory Model. 22 (2018) 394–409.

[20]    W. Han, Z. Chen, Effects of finite-rate droplet evaporation on the extinction of spherical burner-stabilized diffusion flames, Int. J. Heat Mass Transf. 99 (2016) 691–701.

[21]    Y. Zhuang, H. Zhang, Effects of water droplet evaporation on initiation, propagation and extinction of premixed spherical flames, Int. J. Multiph. Flow. 117 (2019) 114–129.

[22]    G. Joulin, P. Clavin, Linear stability analysis of nonadiabatic flames: Diffusional-thermal model, Combust. Flame. 35 (1979) 139–153.

[23]    P. Clavin, Dynamic behavior of premixed flame fronts in laminar and turbulent flows, Prog. Energy Combust. Sci. 11 (1985) 1–59.

[24]    Z. Chen, Y. Ju, Theoretical analysis of the evolution from ignition kernel to flame ball and planar flame, Combust. Theory Model. 11 (2007) 427–453.

[25]    Q. Li, C. Liu, H. Zhang, M. Wang, Z. Chen, Initiation and propagation of spherical premixed flames with inert solid particles, Combust. Theory Model. 24 (2020) 606–631.

[26]    R. Blouquin, P. Cambray, G. Joulin, Radiation-affected dynamics of enclosed spherical flames propagating in particle-laden premixtures, Combust. Sci. Technol. 128 (1997) 231–255.

[27]    R. Blouquin, G. Joulin, Radiation-affected hydrodynamic instability of particle-laden flames, Combust. Sci. Technol. 110–111 (1995) 341–359.

[28]    Y. Ju, C.K. Law, Dynamics and extinction of non-adiabatic particle-laden premixed flames, Proc. Combust. Inst. 28 (2000) 2913–2920.

[29]    T.-H. Lin, C.K. Law, S.H. Chung, Theory of laminar flame propagation in off-stoichiometric dilute sprays, Int. J. Heat Mass Transf. 31 (1988) 1023–1034.

[30] S. Hayashi, S. Kumagai, Flame propagation in fuel droplet-vapor-air mixtures, Symp. (Int.) Combust. 15 (1975) 445–452.
[31] W.E. Ranz, J. W. R. Marshall, Evaporation from Drops, Part I., Chem. Eng. Prog. 48 (1952) 141–146.
[32] L. He, Critical conditions for spherical flame initiation in mixtures with high Lewis numbers, Combust. Theory Model. 4 (2000) 159–172.
[33] H. Zhang, Z. Chen, Spherical flame initiation and propagation with thermally sensitive intermediate kinetics, Combust. Flame. 158 (2011).
[34] H. Zhang, P. Guo, Z. Chen, Outwardly propagating spherical flames with thermally sensitive intermediate kinetics and radiative loss, Combust. Sci. Technol. 185 (2013) 226–248.
[35] H. Zhang, Z. Chen, Effects of heat conduction and radical quenching on premixed stagnation flame stabilised by a wall, Combust. Theory Model. 17 (2013).
[36] M.L. Frankel, G.I. Sivashinsky, On effects due to thermal expansion and Lewis number in spherical flame propagation, Combust. Sci. Technol. 31 (1983) 131–138.
[37] M. Frankel, G. Sivashinsky, On quenching of curved flames, Combust. Sci. Technol. 40 (1984) 257–268.
[38] H. Li, H. Zhang, Z. Chen, Effects of endothermic chain-branching reaction on spherical flame initiation and propagation, Combust. Theory Model. (2018) 1–19.
[39] A.H. Lefebvre, V.G. McDonell, Atomization and sprays, CRC press, 2017.
[40] H. Zhang, P. Guo, Z. Chen, Critical condition for the ignition of reactant mixture by radical deposition, Proc. Combust. Inst. 34 (2013) 3267–3275.
[41] J.W. Dold, Premixed flames modelled with thermally sensitive intermediate branching kinetics, Combust. Theory Model. 11 (2007) 909–948.
[42] M. Abramowitz, I.A. Stegun, R.H. Romer, Handbook of mathematical functions with formulas, graphs, and mathematical tables, (1988).
[43] R.K.S. Hankin, Special functions in R: introducing the gsl package, Newsl. R Proj. Vol. 6/4, Oct. 2006. 6 (2006) 24.
[44] W. Han, Z. Chen, Effects of finite-rate droplet evaporation on the ignition and propagation of premixed spherical spray flame, Combust. Flame. 162 (2015) 2128–2139.
[45] Z. Chen, Y. Ju, Theoretical analysis of the evolution from ignition kernel to flame ball and planar flame, Combust. Theory Model. 11 (2007) 427–453.
[46] B. Deshaies, G. Joulin, On the initiation of a spherical flame kernel, Combust. Sci. Technol. 37 (1984) 99–116.
[47] C.K. Law, Combustion Physics, Cambridge University Press, New York, 2006.
[48] J.B. Greenberg, A. Kalma, A study of stretch in premixed spray flames, Combust. Flame. 123 (2000) 421–429.
[49] M. Mortensen, R.W. Bilger, Derivation of the conditional moment closure equations for spray combustion, Combust. Flame. 156 (2009) 62–72.
[50] S. Ukai, A. Kronenburg, O.T. Stein, LES-CMC of a dilute acetone spray flame, Proc. Combust. Inst. 34 (2013) 1643–1650.
[51] A. Giusti, E. Mastorakos, Detailed chemistry LES/CMC simulation of a swirling ethanol spray flame approaching blow-off, Proc. Combust. Inst. 36 (2017) 2625–2632.
[52] A.Y. Klimenko, R.W. Bilger, Conditional moment closure for turbulent combustion, Prog. Energy Combust. Sci. 25 (1999) 595–687.

[53] M. Bolla, Y.M. Wright, K. Boulouchos, G. Borghesi, E. Mastorakos, Soot formation modeling of n-heptane sprays under diesel engine conditions using the conditional moment closure approach, Combust. Sci. Technol. 185 (2013) 766–793.

[54] A. Wehrfritz, O. Kaario, V. Vuorinen, B. Somers, Large eddy simulation of n-dodecane spray flames using flamelet generated manifolds, Combust. Flame. 167 (2016) 113–131.

[55] G.I. Barenblatt, V.B. Librovich, G.M. Makhviladze, The mathematical theory of combustion and explosions, New York Consult. Bur. (1985).